\documentclass[a4paper,10pt]{article}
\usepackage{amsmath,amssymb,graphics,graphicx,dcolumn,bm,enumerate}
\usepackage{color}
\usepackage[affil-it]{authblk}
\usepackage{textpos}

\usepackage[paperheight=30cm,left=2cm,right=2cm,top=1cm, bottom=2cm]{geometry}

\begin{document}
 \title{Quantum Annealing and  Computation: A Brief Documentary Note\footnote{Dedicated to Prof. Bikas K. Chakrabarti on the occasion of his 60-th birthyear}}

\author{Asim Ghosh%
  \thanks{Electronic address: \texttt{asim.ghosh@saha.ac.in}}}
\author{Sudip Mukherjee%
  \thanks{Electronic address: \texttt{sudip.mukherjee@saha.ac.in}}}
\affil{Condensed Matter Physics Division \\ Saha Institute of Nuclear Physics \\ 1/AF Bidhannagar, Kolkata 700 064, India.}

\maketitle
\begin{textblock*}{3cm}(5cm,-7cm)
      \fbox{\footnotesize SCIENCE AND CULTURE (Indian Science News Association), vol. \textbf{79}, pp. 485-500 (2013)}
    \end{textblock*}

\begin{abstract}
 \noindent Major breakthrough in quantum computation has recently been achieved using quantum annealing to develop analog quantum computers instead of gate based  computers. After a short introduction to quantum computation,   we retrace very briefly  the history of these developments and discuss the Indian researches  in this connection and provide some interesting documents (in the Figs.) obtained from a chosen set of high impact papers (and also some recent news etc. blogs appearing in the Internet). This note is also designed to supplement an earlier note by Bose (Science and  Culture, \textbf{79}, pp. 337-378, 2013).                                                                                                                                                                                                                                                                                                                                      
\end{abstract}
\vskip 0.5cm

\section{Introduction}\vskip 0.5cm
Quantum computers are actively being   sought for the last couple of decades. Basic hope being that  quantum mechanics promises several features to help faster computations if quantum features are properly implemented in the hardware architecture of such computers. Traditional  architecture of   classical computers is logical gate-based ones. The linear superposition processing  of the wave functions in quantum mechanics helps simultaneous (probabilistic) processing of the binary bits or qubits (of information). Quantum mechanics also promises   major advantages of  parallel operation of these gates in appropriate architectures. The problem of  decoherence  has not allowed so far any gate-based quantum computer   which is able to handle more than a couple of qubits. Even in classical computers, in order to solve computationally hard problems, like the traveling salesman  problem, one artificially generates stochastic algorithms, like the simulated annealing techniques (in the so-called `Boltzmann Machine') for practical and efficient searches. The essential  problem in such searches  is that the system gets locked  in some local minimum separated  from the other deeper minimum by (cost function or free energy) barriers. Noting that quantum tunneling feature across such barriers can help \cite{ray-89} searching for the minimum (optimal solution), quantum annealing techniques have been  developed recently [1-29] (see Figs. 1-28). In the last couple of years, such techniques have  been efficiently  implemented in the computer architectures and such quantum annealing computers have already arrived in the market (see Fig. \ref{bbc.pdf}) and major successes are being demonstrated \cite{johnson-11,smelyanskiy-12, Nagaj-12, Titiloye-12,Yamamoto-12,Ortiz-12,Bapst,Boixo-nature-13,Boixo-13}. These exciting developments are also being captured in several recent notes (e.g., \cite{bose-13}), reviews (e.g., \cite{Bapst,suzuki-13}) and books (e.g., \cite{Dutta-book,bkc-book}). 

\section{A brief history}\vskip 0.5cm
Ray et al. \cite{ray-89} and Thirumalai et al.  \cite{Thirumalai-89} suggested (in 1989; see Figs. \ref{ray-89.pdf} and \ref{thirumalai-89.pdf}) that quantum tunneling from local trap states across the (narrow) free energy  barriers of macroscopic height (of order $N$; coming from the requirement of the flips of finite fraction of all $N$ spins) in a Sherrington-Kirppatrick spin glass (in transverse  field) can help evolving such a complex system towards its (degenerate) ground state(s). Although the idea was criticized heavily in the subsequent literature, essentially on the ground that the incoherent phase overlaps of the tunneling states (waves) will localized the system and will not allow evolution towards the  ground state(s).  Later, theoretical investigations (\cite{finnila-94,kadowaki,farhi-01,Santoro-02}) and the experimental demonstrations (\cite{brooke-99}) lead to a very promising development resulting in the quantum annealing technique \cite{book-das,santoro-06,das-08,morita-08,Yamamoto-12} with the hardware implementation by D-Wave system \cite{johnson-11}. Successful checking  and applications \cite{smelyanskiy-12,Ortiz-12,Boixo-nature-13,Boixo-13} led to the emergence of a new era in quantum computing  (see e.g., the comments by Bose \cite{bose-13}, Fig. \ref{bose-13.pdf}). 
\vskip 1cm
{\color{white}.}
\vskip 1.5cm
\section{A short story of the development}
\begin{figure}[!htbp]
\centering
\includegraphics[width=15cm]{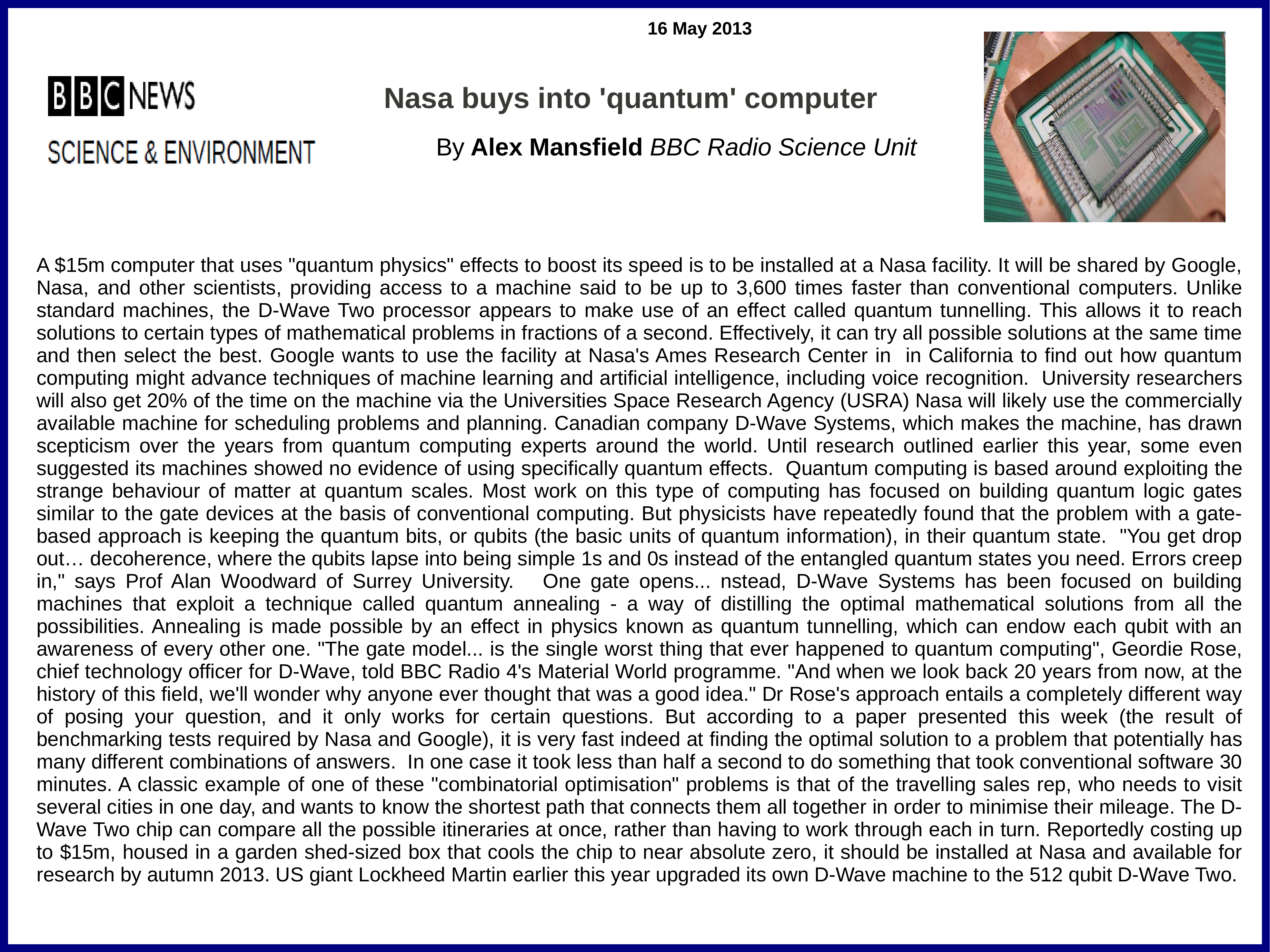}
\caption{BBC news blog on the purchase deal of the latest 512 qubit quantum annealer of D-Wave Inc. by the NASA-Google consortium (website: http://www.bbc.co.uk/news/science-environment-22554494).}
\label{bbc.pdf}
\end{figure}

\begin{figure}[!htbp]

\includegraphics[width=15cm]{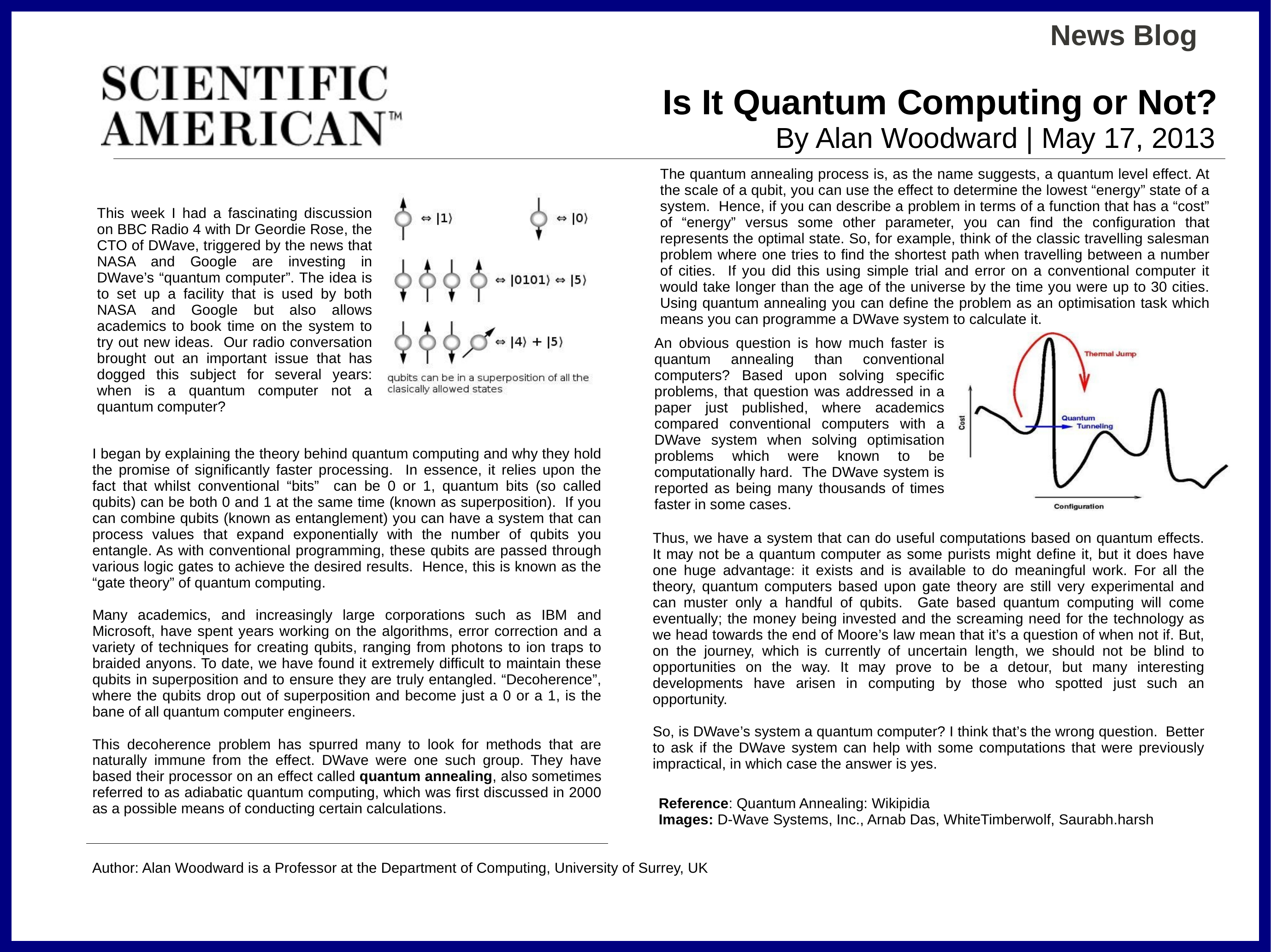}
\centering
\caption{A Scientific American news blog on the NASA-Google  investment in D-Wave quantum (annealing) computer. It explains briefly the working principle involved, using the Wikipedia entry on quantum annealing (Fig. \ref{wiki.pdf}; website: http://blogs.scientificamerican.com/guest-blog/2013/05/17/is-it-quantum-computing-or-not/). }
\label{sci-ame.pdf}
\end{figure}

\begin{figure}[!htbp]
\centering
\includegraphics[width=17cm]{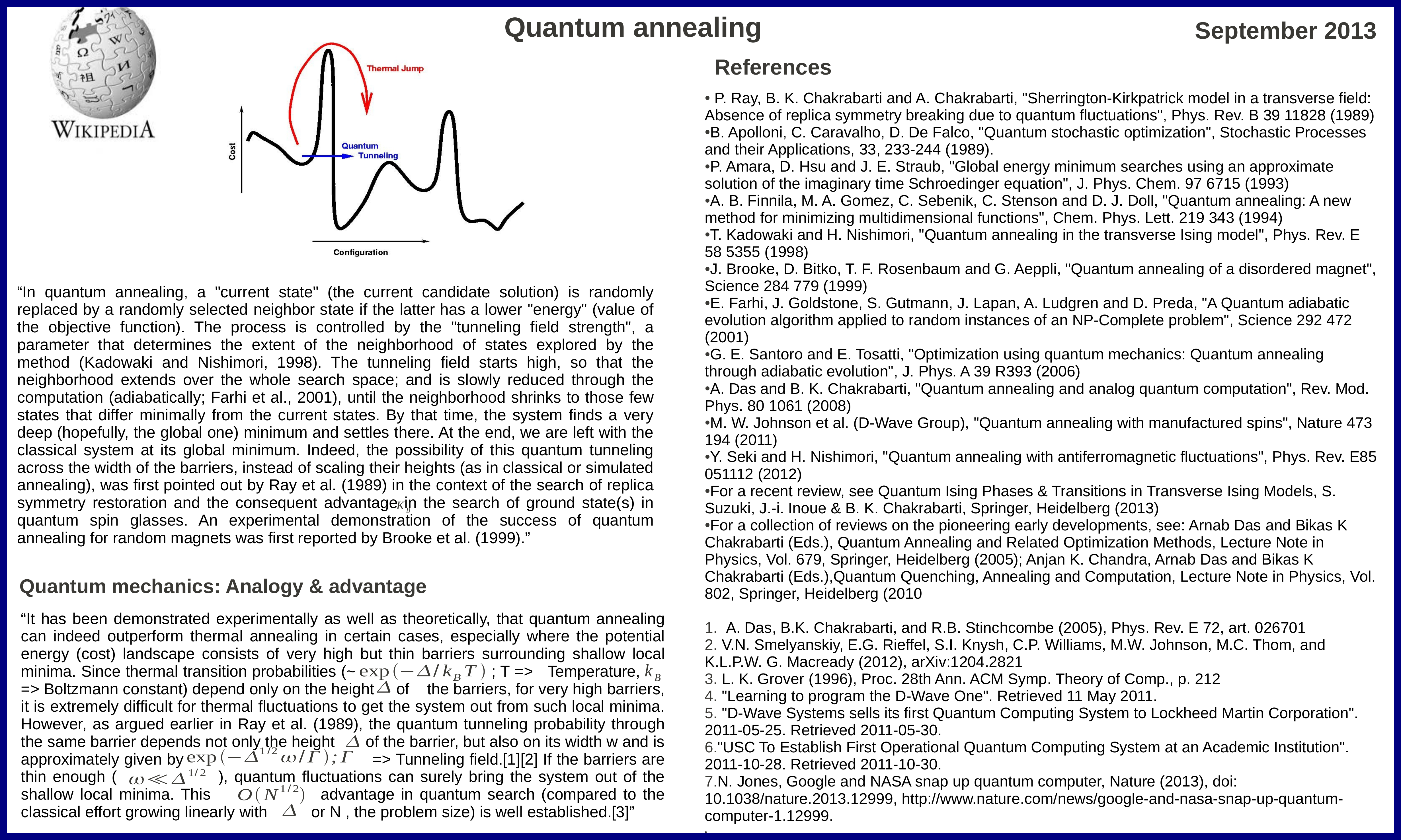}
\caption{Part of the entry on quantum annealing in Wikipedia (as in September 2013; website: http://en.wikipedia.org/wiki/Quantum-annealing ).}
\label{wiki.pdf}
\end{figure}

\vskip 0.5cm
In May this year (2013), there have been several news posts and blogs  in popular science journals as well
as newspapers informing  about the successful tests of a quantum computer with about 100 qubits (order of
magnitude higher than those available otherwise), based on quantum annealing technique and marketed by the D-Wave
Systems Inc. Interestingly, the NASA group of Consortium had already placed an   order to them  for a 512 qubit
quantum annealing computer. Fig.  \ref{bbc.pdf}  shows the BBC news blog on the purchase deal with D-Wave Inc. by the NASA-Google consortium. In an attempt to explain in a popular way how  quantum  tunneling can help such analog
computers to get out of the ``local'' solutions and anneal down quantum mechanically to the ``global''  or ``optimal'' solutions with proper tuning of the tunneling term, a Scientific American news blog appeared last May (see Fig. \ref{sci-ame.pdf}). It explains in brief the working principle involved, using the Wikipedia entry on quantum annealing (partly reproduced in Fig.  \ref{wiki.pdf}).

\begin{figure}[!htbp]\centering

\includegraphics[width=15cm]{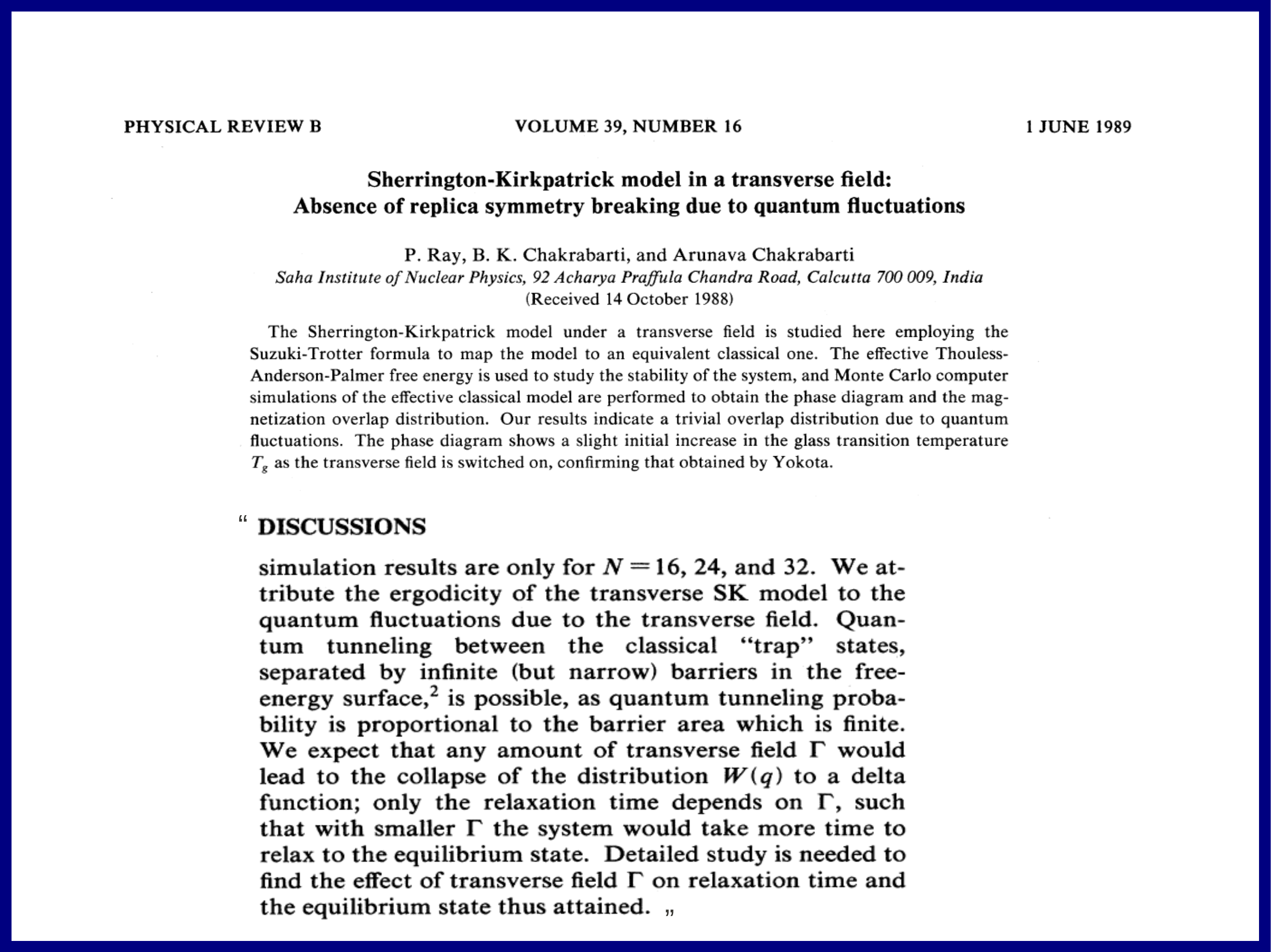}
\caption{Title, abstract and some excerpts from the first published paper arguing that quantum tunneling across the free energy barriers in Sherrienton-Kirkpatrick spin glass model can lead to efficient search possibilities for its ground state(s). It may be noted that computationally hard problems can often be mapped into such long-range spin glass models and the advantage of quantum tunneling in such quantum spin glass models has lead  ultimately to the development of the quantum annealer/computer discussed here.}
\label{ray-89.pdf}
\end{figure}

\begin{figure}[!htbp]\centering
\includegraphics[width=12cm]{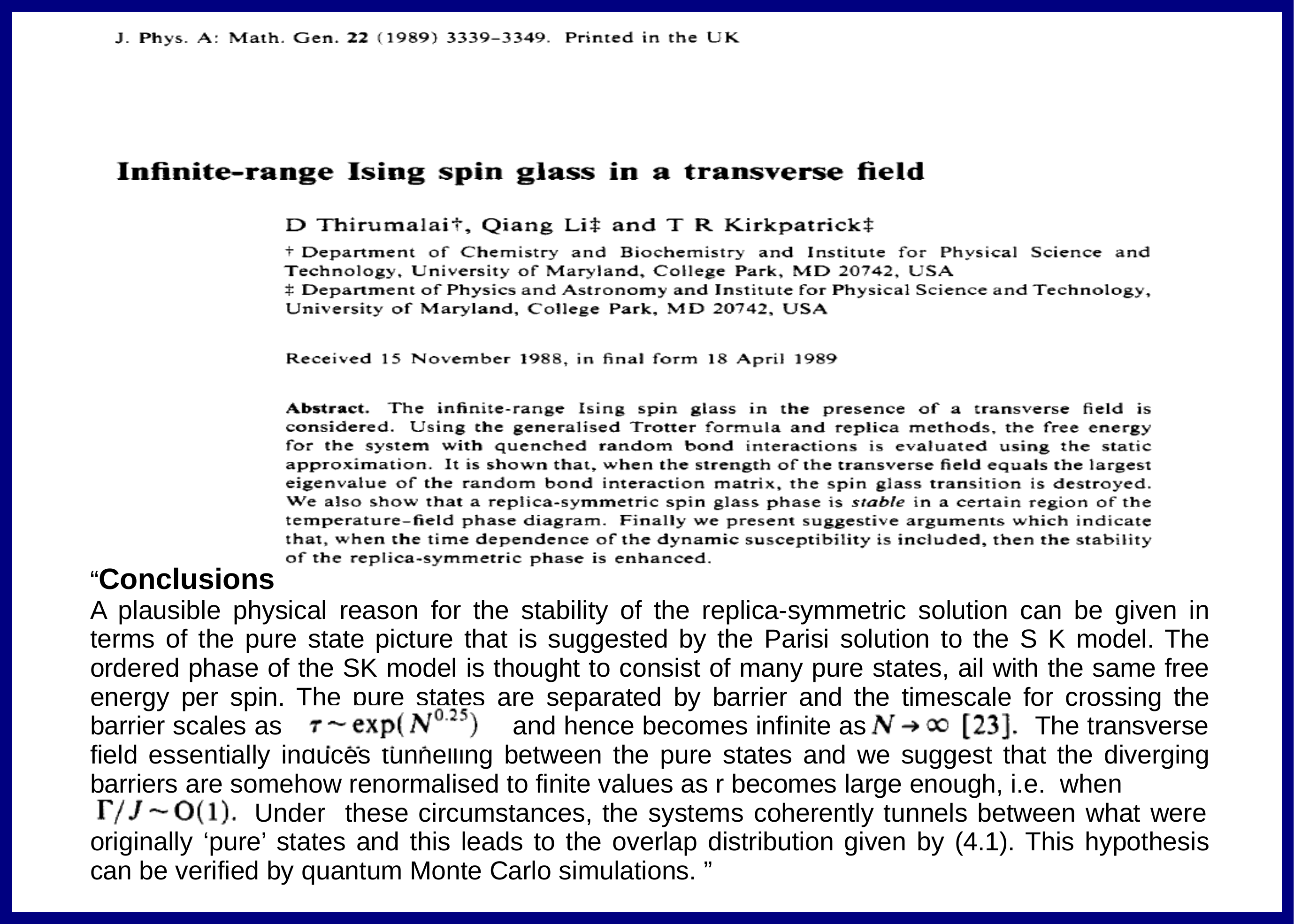}
\caption{Title, abstract and some excerpts from \cite{Thirumalai-89}, arguing (in almost the same language as in ref. \cite{ray-89}; see Fig. \ref{ray-89.pdf}) for the possible advantage of quantum tunneling between well characterised (classically) localized states in the same long-range transverse Ising spin glass model.}
\label{thirumalai-89.pdf}
\end{figure}

The rugged nature of the (free) energy landscape (energy versus spin configurations) of a classical (Ising) spin
glass does not allow searches for the global energy minima by simple rolling down the landscape (say, using an energy dissipative dynamics). Essentially the system gets trapped in the local minima, separated from the global minima often by macroscopically high (free) energy barriers. Ray et al. \cite{ray-89} first pointed out that if such barriers are narrow, quantum mechanical tunneling (as in a transverse Ising spin glass model) can help such searches (see Fig.  \ref{ray-89.pdf}).  As mentioned already, this indication was criticized heavily in the
following literatures. However, some crucial features of such tunneling effects were immediately checked, with
positive results in some solid-state samples discovered by Wu  et al. (see Fig. \ref{wu-91.pdf}). The possibility of tuning the quantum tunneling term to achieve the minimization of a multi-variable optimization problem was pointed out by Finnila et al. \cite{finnila-94} (see Fig. \ref{finnila-94.pdf}). However, the use of this annealing (of the
tunneling field) in the well-characterized ground state search problems of (frustrated) spin systems were convincingly demonstrated, using numerical techniques, by Kadwaki and Nishimori \cite{kadowaki} (see also Fig. \ref{nishimori-98.pdf}) and the reported success in this paper made a major impact on the following developments. Soon, Brook et al. \cite{brooke-99} extended their earlier experimental investigations (see Fig.  \ref{wu-91.pdf}) and with suitable tuning of the tunneling field, observed  clear advantages of the quantum annealing in the search for ground state(s) for such samples (see Fig.  \ref{aeppli-99.pdf}). This experimental demonstration of the clear advantages of quantum annealing established the field. Soon major investigations by several groups from all over the world started. Farhi et al. \cite{farhi-01} (see Fig. \ref{farhi-01.pdf}) indicated that such an adiabatic evolution (zero temperature annealing) algorithm may help solving the computationally (NP-) hard problems. The estimate of the growth of errors in the optimal search, with the decrease in the annealing time, was made  first by Santoro et al. \cite{Santoro-02} (see Fig.  \ref{santoro-02.pdf}) and the extensions (and some clarifications) of the experimental investigations by Ancona-Torres et al. (see Fig.  \ref{aeppli-PRL-2008.pdf}) enthused the community to explore further in much more meaningful way. Soon (during 2006-2008), some important reviews on quantum annealing techniques were written (see Figs. \ref{Santoro-06.pdf}, \ref{das-rmp.pdf} and \ref{Morita-08.pdf}): Santoro and Tosatti \cite{santoro-06} and Morita and Nishimori \cite{morita-08} reviewd the mathematical
structures of the quantum annealing techniques, while Das and Chakrabarti \cite{das-08} reviewed the physical structure of quantum annealing and discussed its possible implications for analog quantum  computations.\footnote{For extensions to non-adiabatic quantum annealing see M. Ohzeki, and H. Nishimori, 
J. Comp. and Theor. Nanoscience \textbf{8}, 963(2011) and  D. Aharonov et al., SIAM J.  Comp., \textbf{37},  166 (2007),  http://arxiv.org/abs/quant-ph/0405098, for  showing complexity equivalence between adiabatic quantum computation and circuit computation. \vskip -0.1cm}

In 2011, D-wave system announced \cite{johnson-11} the arrival of ``World's first commercially available quantum computer, operating on a 128 qubit chip-set using quantum annealing'' (see Figs. \ref{D-wave.pdf}, \ref{Johnson-12.pdf}). As indicated already, this news created huge enthusiasms as well as a lot of criticisms
in the community of scientists. However, soon some leading research groups came forward to test the performances of
these machines with remarkably positive results (see e.g., Figs. \ref{ortiz-12.pdf}, \ref{arXiv.pdf}, \ref{boixo-12.pdf} and  \ref{bioxo-13.pdf}). There were parallel investigations on the possible performances of the quantum annealing technique in the context of various computationally hard problems (see e.g., Figs.  \ref{yamamoto.pdf}, \ref{Somma-12.pdf} and  \ref{titiloye-12.pdf}).\footnote{\noindent Of course, the story is not a closed one and many critical aspects of the development are also being addressed by
the scientists. See for example:

\noindent B. Altshuler et al., Proc. Nat. Acad. Sc., \textbf{107},  12446 (2010),
http://www.pnas.org/content/107/28/12446.full
(on why it may fail for general  disorder);

\noindent C. R. Laumann et al., Phys. Rev. Lett. \textbf{109}, 030502 (2012),  http://arxiv.org/abs/1202.3646
(showing both unexpected success and failure modes of
quantum annealing);

\noindent E. Farhi et al.,  Phys. Rev. A \textbf{86}, 052334 (2012), http://arxiv.org/abs/1208.3757
(discussing how Quantum Annealing  can fail in some
generic satisfiability problems); 
}



\begin{figure}[!htbp]\centering
\includegraphics[width=12cm]{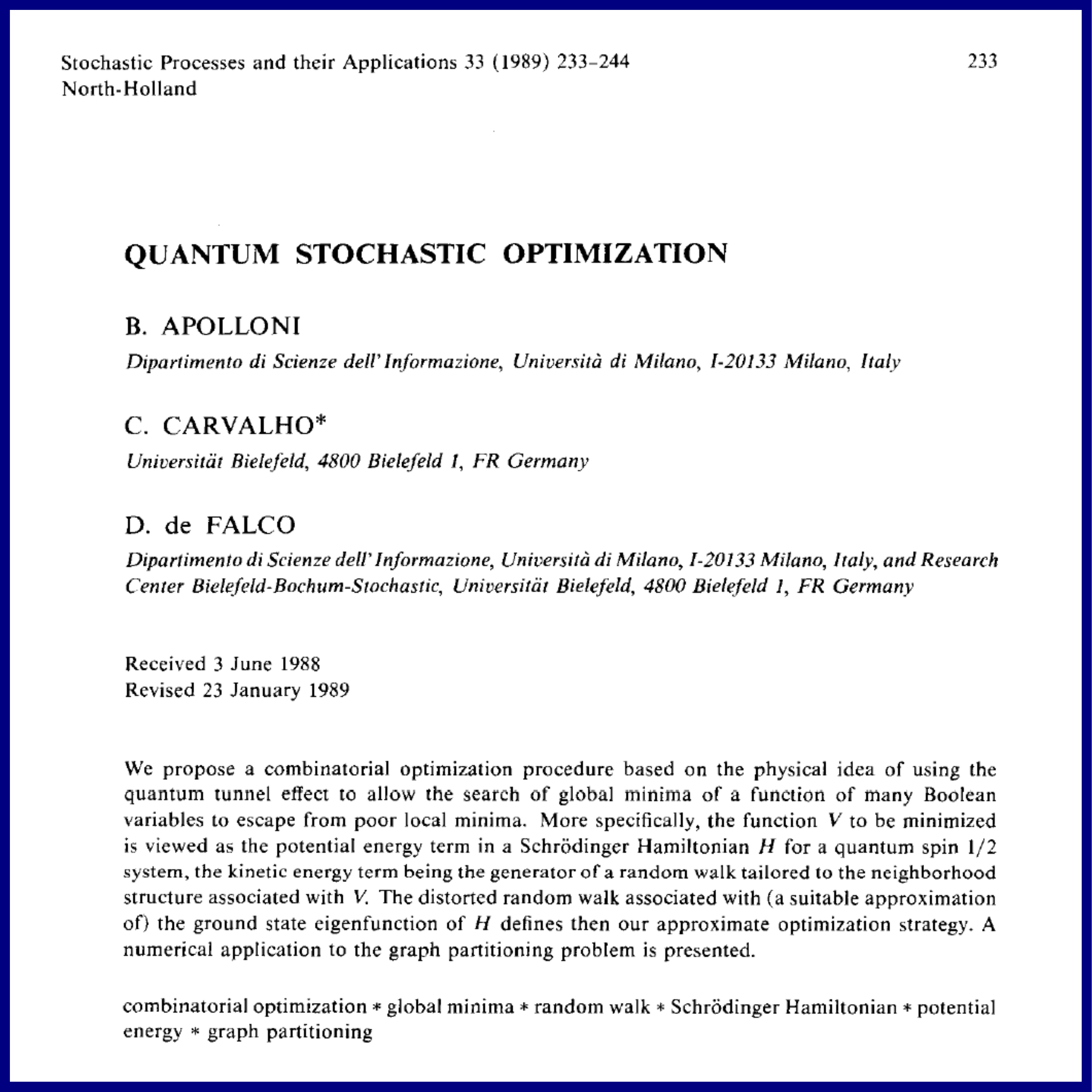}
\caption{Title and  abstract of ref. \cite{apolloni-89} indicating the formulation of a quantum stochastic optimization trick.}
\label{}
\end{figure}

\begin{figure}[!htbp]\centering

\includegraphics[width=15cm]{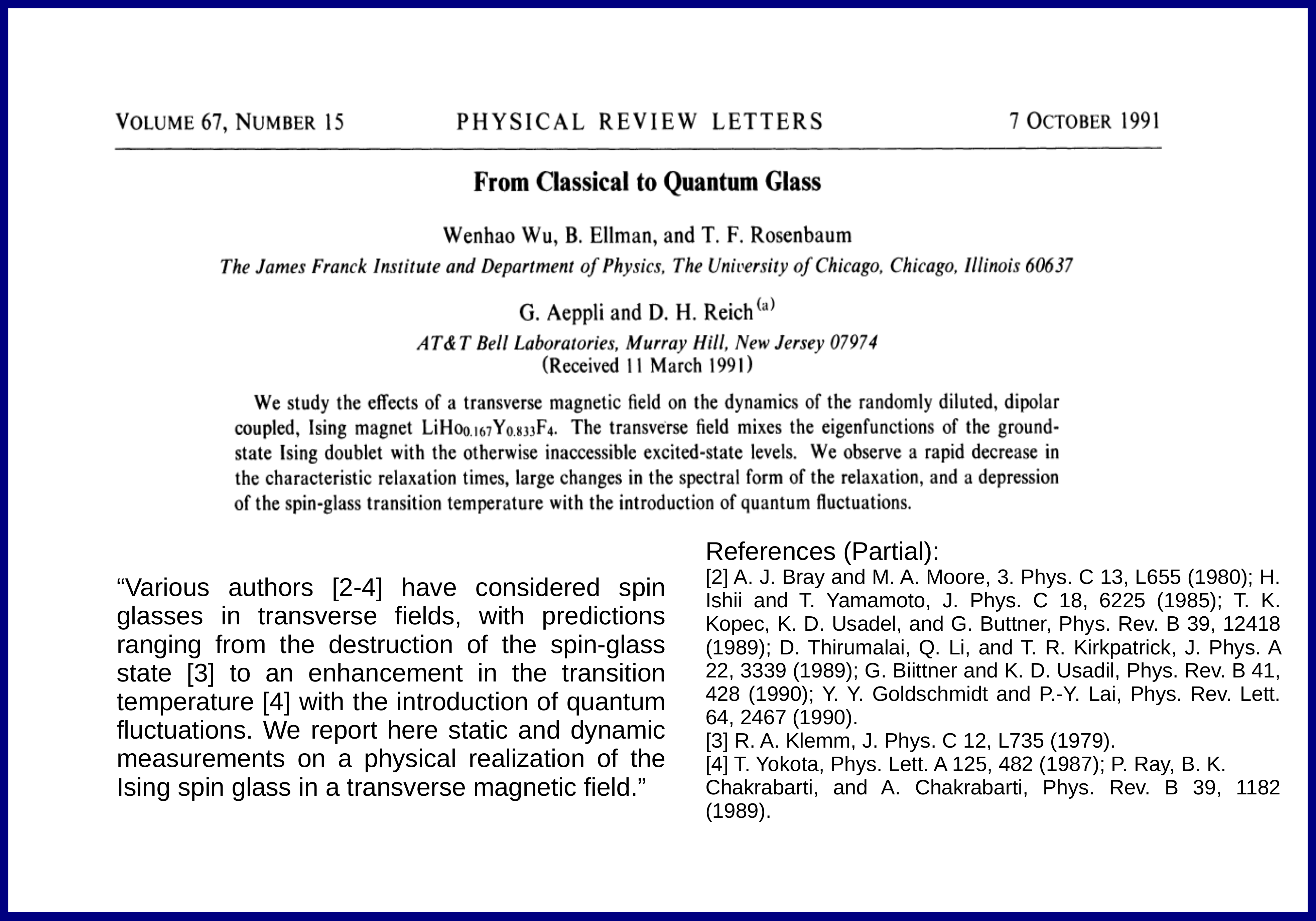}
\caption{Title, abstract and some excerpts from the first paper (from Univ. Chicago and Bell Labs. groups) reporting on the  experimental realization of a sample described precisely by a transverse Ising spin glass model, with tunable  transverse field, and observations  in agreement with the results  of Ray et al. (1989).}
\label{wu-91.pdf}
\end{figure}

\begin{figure}[!htbp]\centering

\includegraphics[width=15cm]{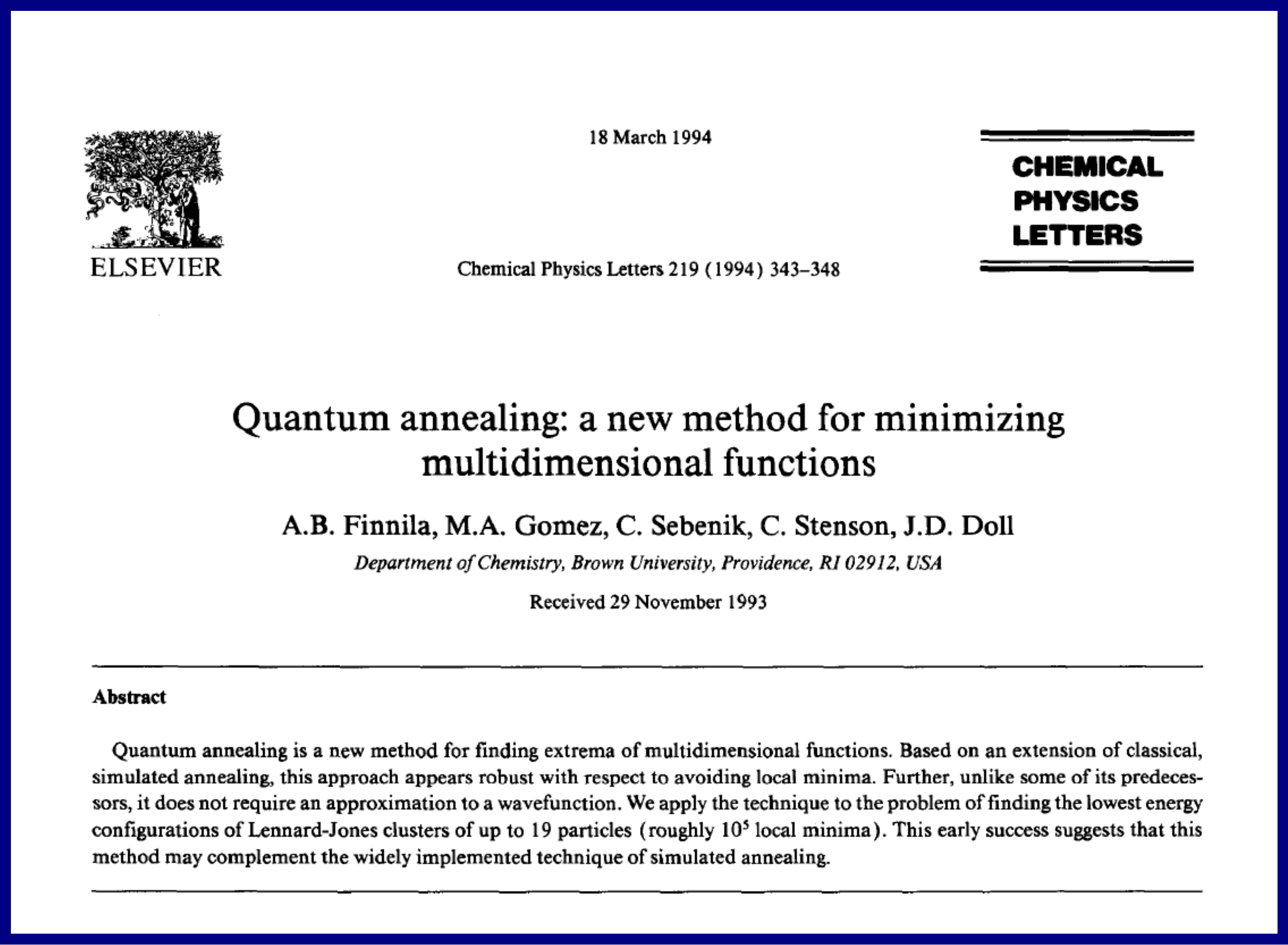}
\caption{Title and  abstract of the first published paper demonstrating the ground state cluster search for a Lennard-Jones system  with `quantum annealing' [in the title; Of course, the first claim for `A numerical implementation of quantum annealing' in a single particle Hamiltonian for minimizing a real function of Boolean variables was published by B. Apolloni, C. Carvalho and D. De Falco in a conference (held in July 1988) proceedings, `Stochastic Processes, Physics \& Geometry', Eds. S. Albeverio et al., World Scientific, Singapore, 1990, pp. 97-111].}
\label{finnila-94.pdf}
\end{figure}

\begin{figure}[!htbp]\centering

\includegraphics[width=15cm]{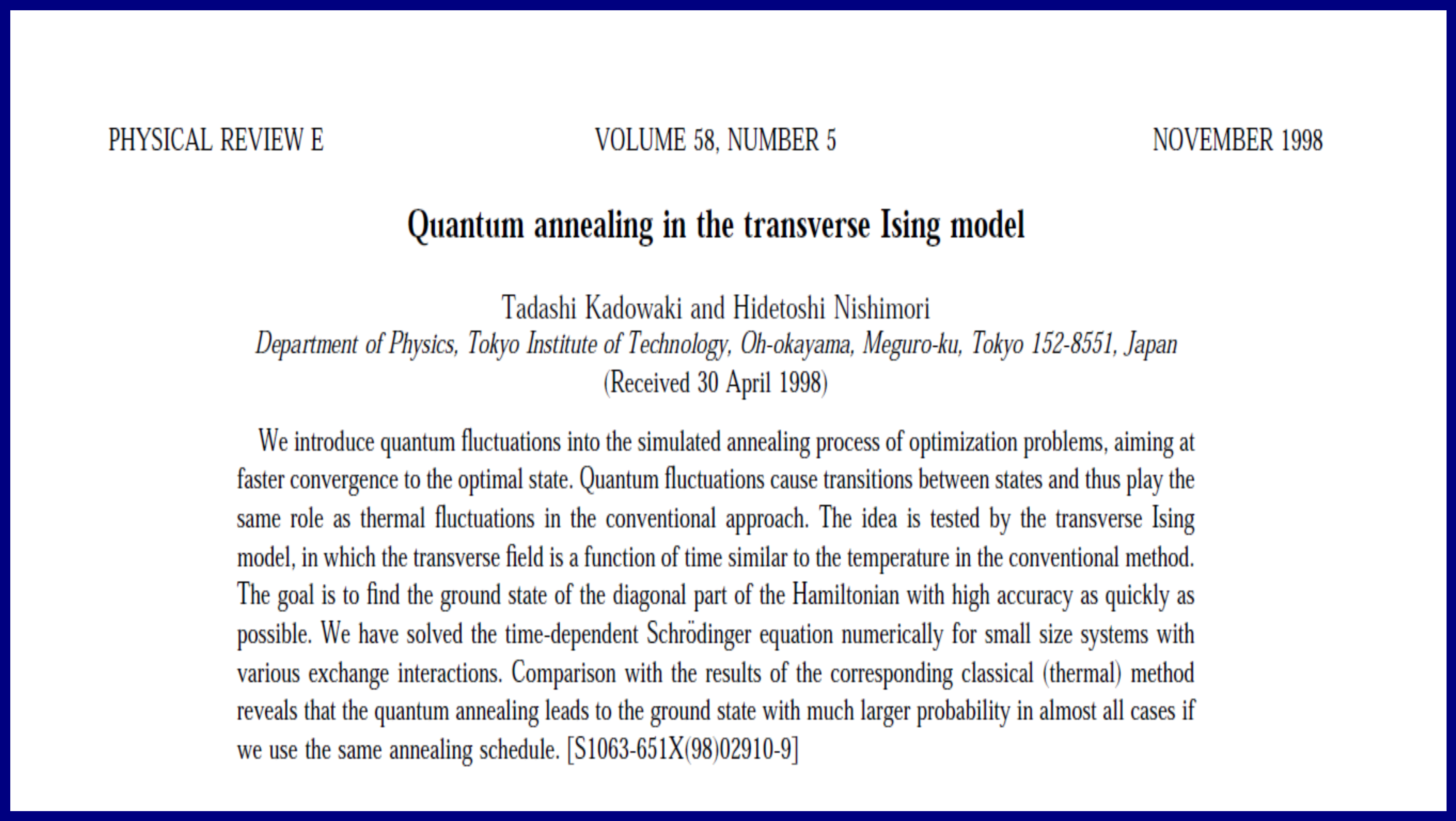}
\caption{Title and  abstract of a   quantum annealing paper demonstrating clear advantages of quantum annealing in well characterized computationally hard problems of Ising models with frustrating interactions. The demonstrations of clear advantages in some well studied spin models made major impact on the subsequent developments.}
\label{nishimori-98.pdf}
\end{figure}

\begin{figure}[!htbp]\centering

\includegraphics[width=15cm]{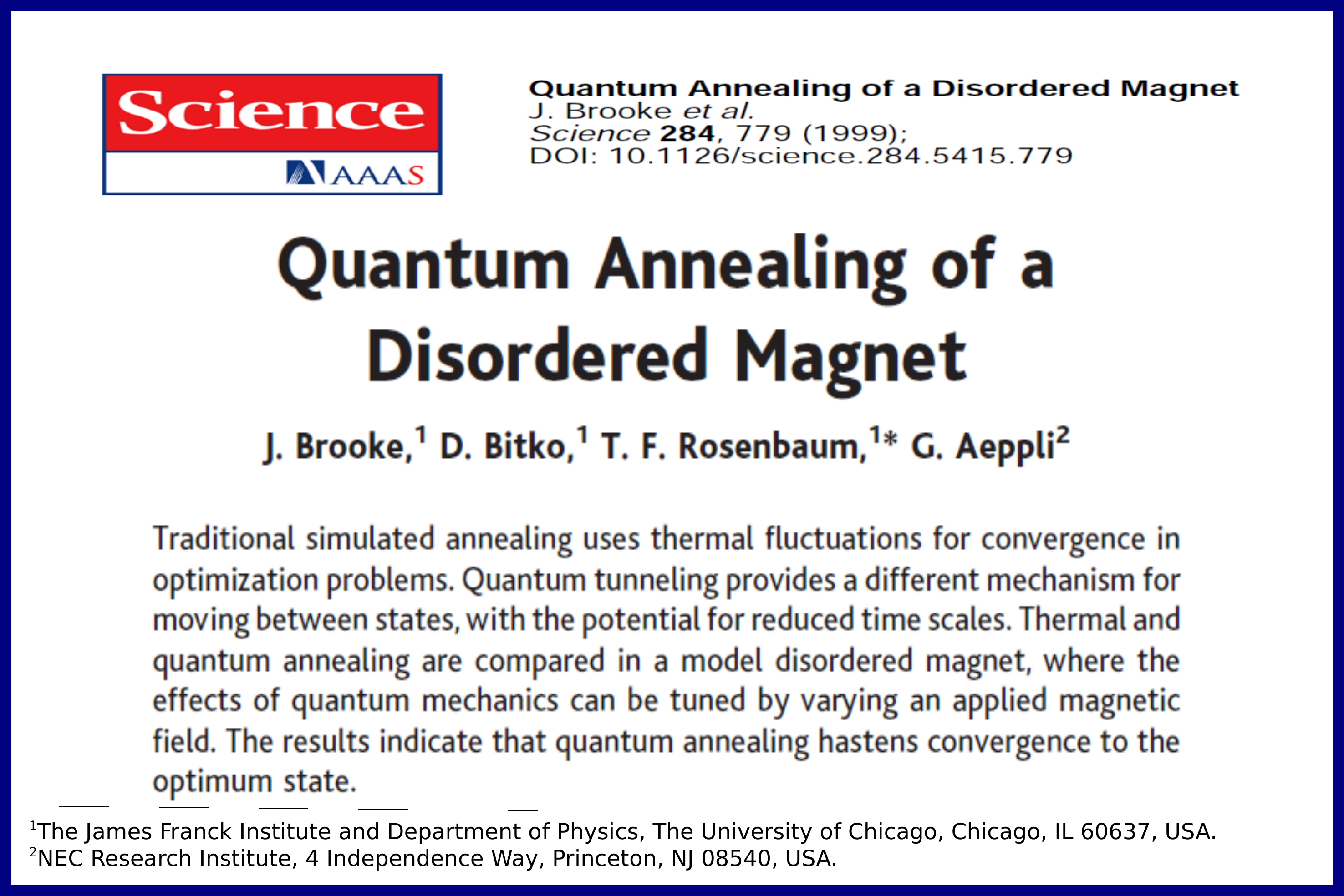}
\caption{Title and  abstract of the first experimental demonstration of the advantages of quantum annealing in extracting ground state of disordered magnets. This experimental demonstration had put the quantum annealing trick on firm physical ground.}
\label{aeppli-99.pdf}
\end{figure}

\begin{figure}[!htbp]\centering

\includegraphics[width=15cm]{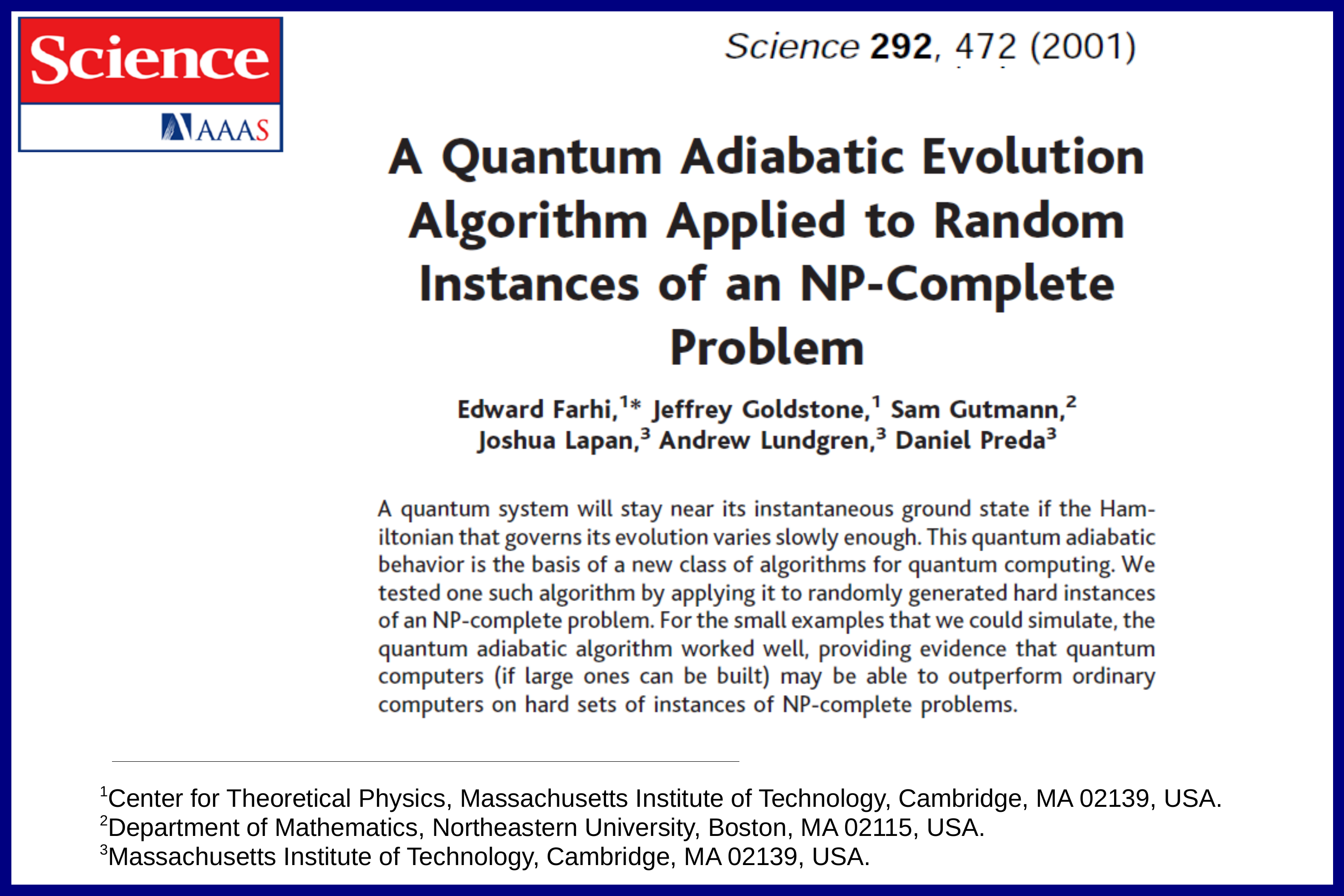}
\caption{Title and abstract of zero temperature quantum annealing algorithm for NP-hard problems.}
\label{farhi-01.pdf}
\end{figure}

\begin{figure}[!htbp]\centering
\includegraphics[width=12cm]{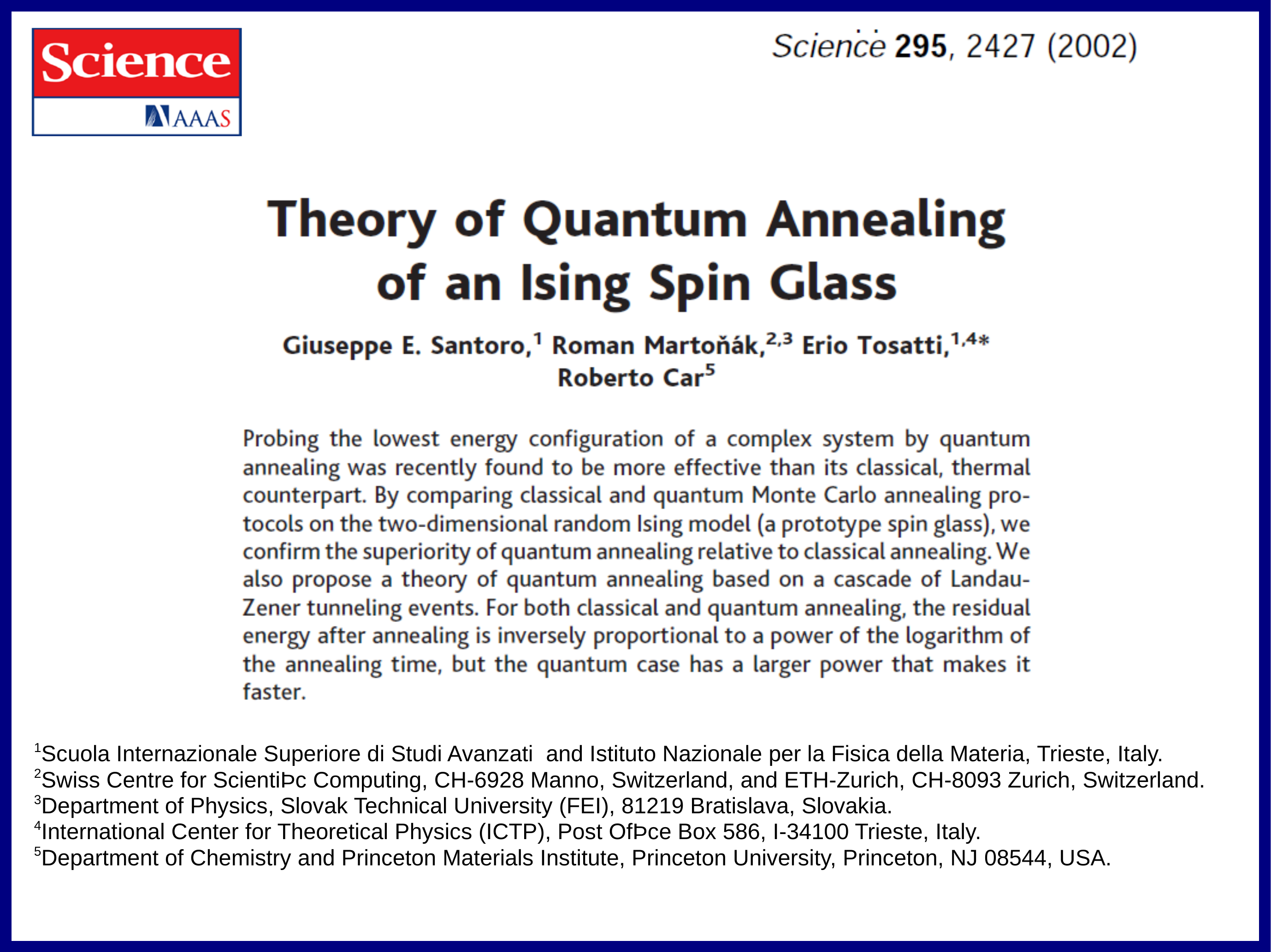}
\caption{Title and abstract of a paper on application of quantum annealing in estimating the remaining fraction of undesired solutions in some optimization searches in Ising spin-glasses.}
\label{santoro-02.pdf}
\end{figure}

\begin{figure}[!htbp]\centering
\includegraphics[width=12cm]{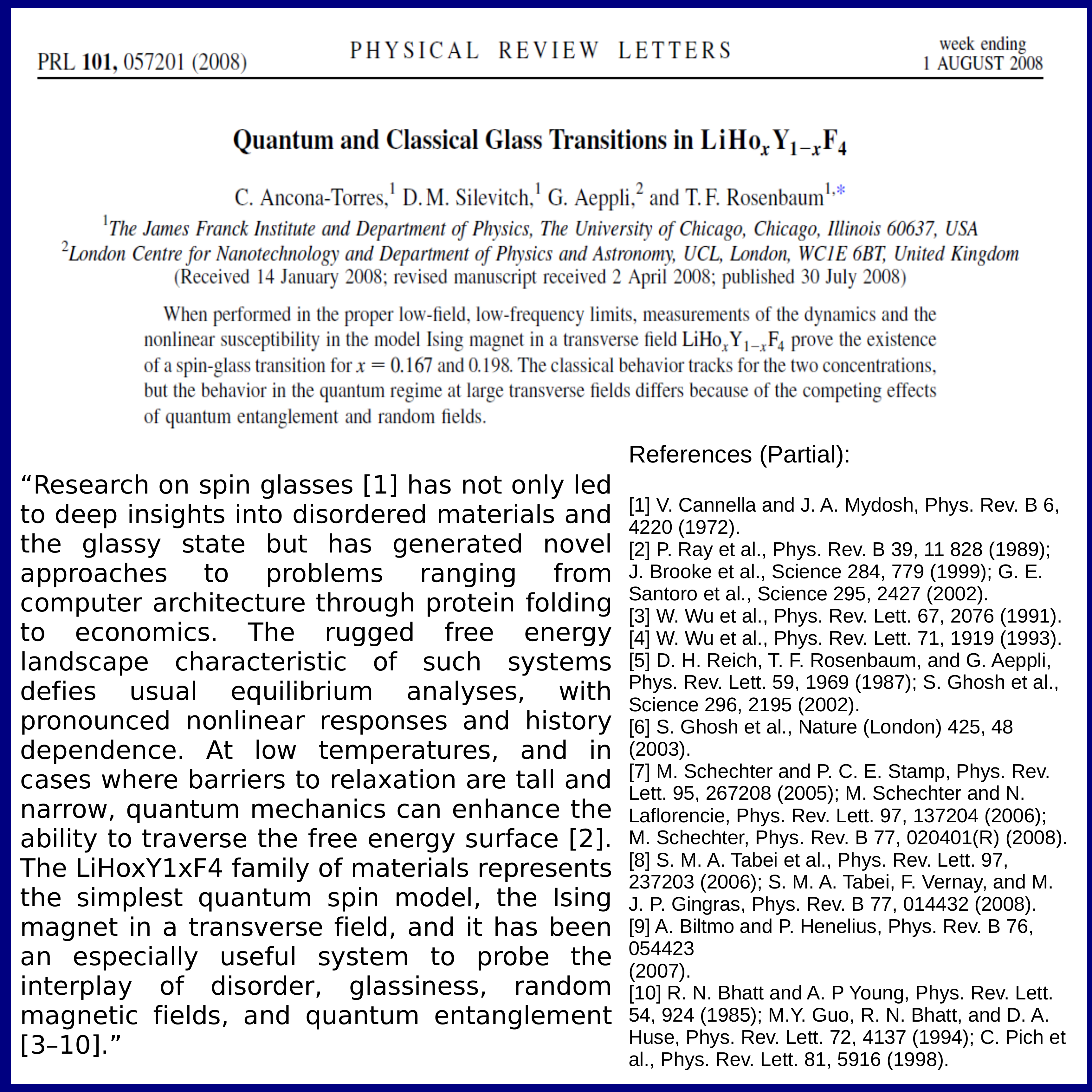}
\caption{Title, abstract and some excerpts from a paper extending and clarifying the method used in ref. \cite{brooke-99} (see fig. \ref{aeppli-99.pdf}) for quantum glasses. }
\label{aeppli-PRL-2008.pdf}
\end{figure}

\begin{figure}[!htbp]\centering
\includegraphics[width=12cm]{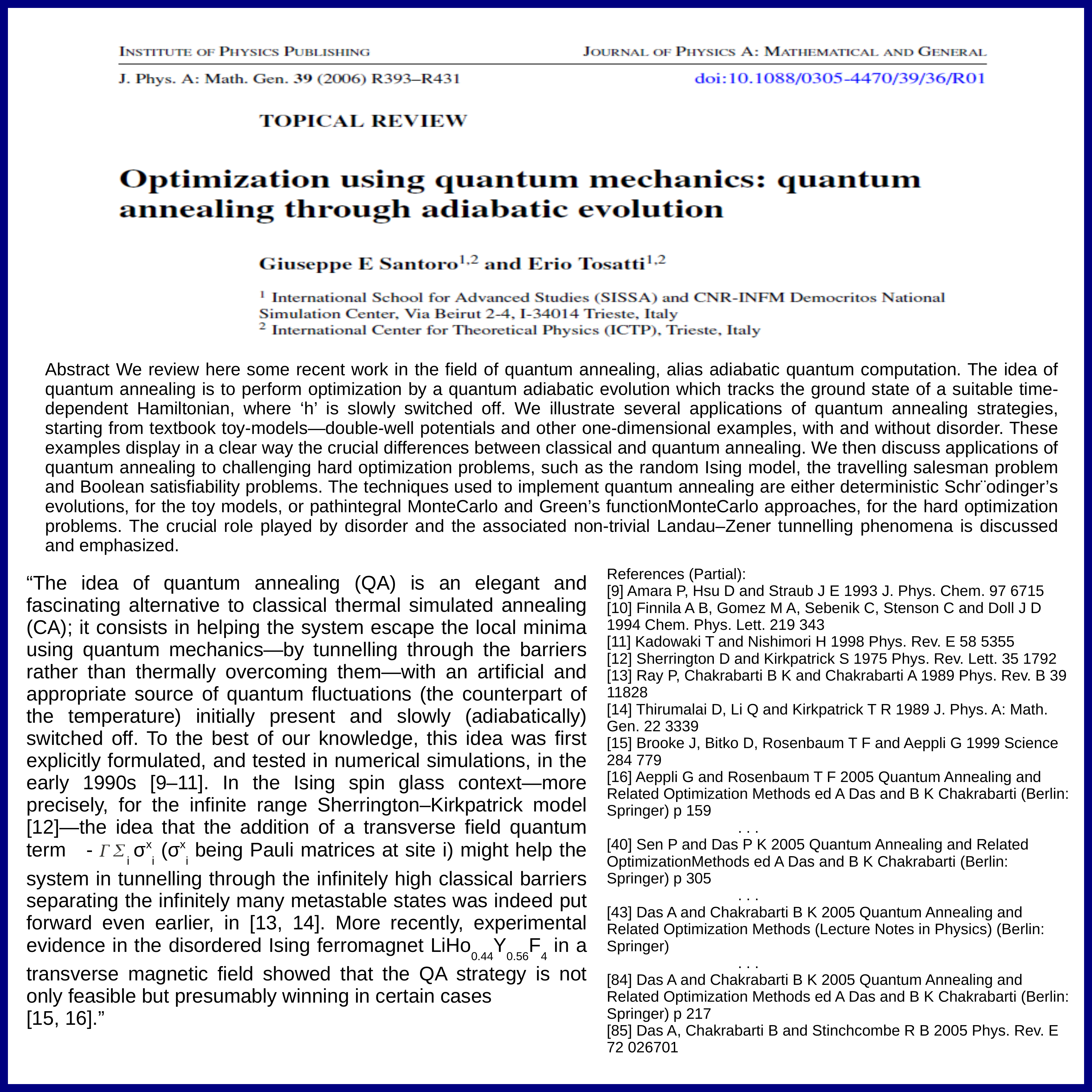}
\caption{Title, abstract and some excerpts for the first review on adiabatic quantum computation/annealing.}
\label{Santoro-06.pdf}
\end{figure}

\begin{figure}[!htbp]\centering

\includegraphics[width=12cm]{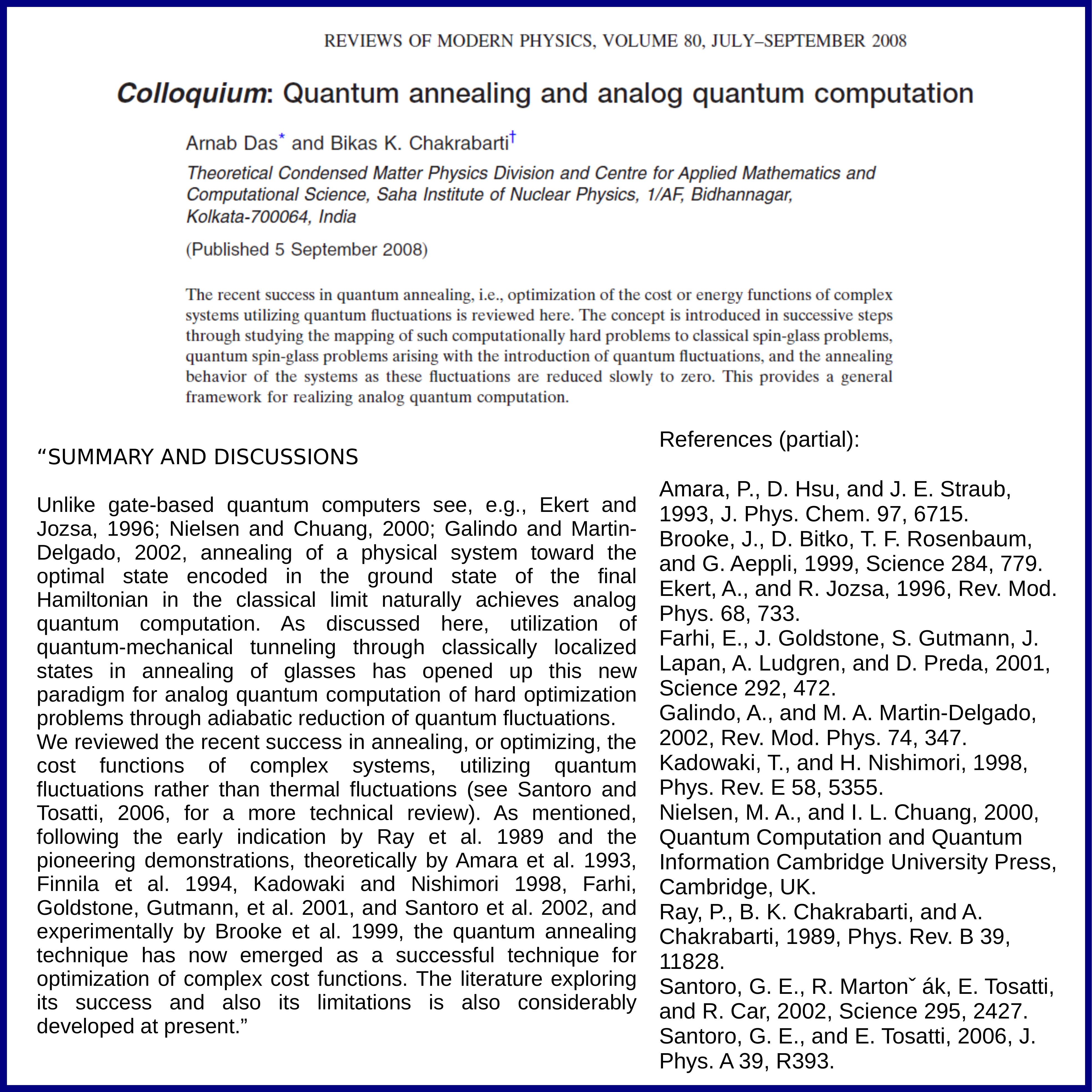}
\caption{Title, abstract and some excerpts from a review on quantum annealing and quantum computation.}
\label{das-rmp.pdf}
\end{figure}

\begin{figure}[!htbp]\centering

\includegraphics[width=12cm]{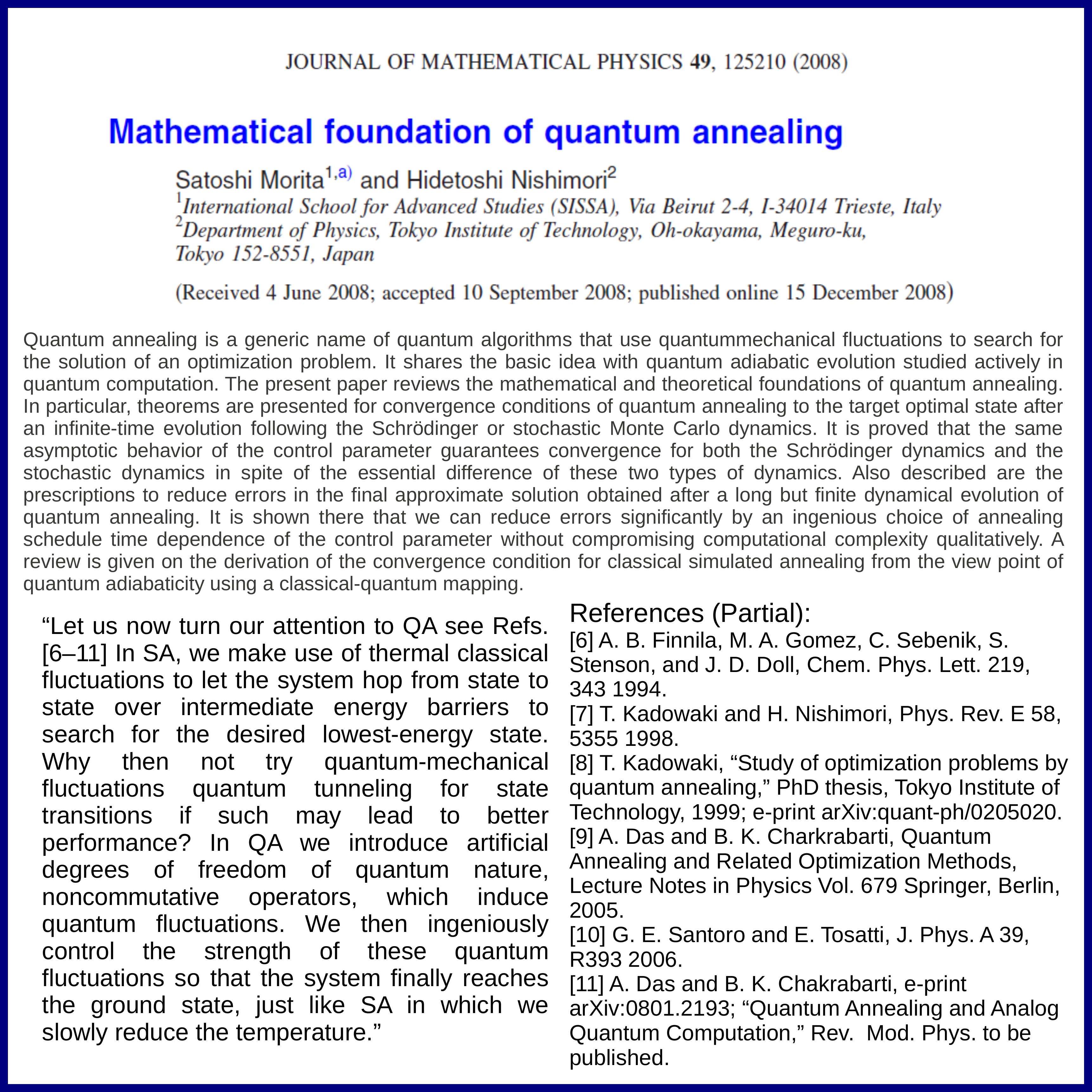}
\caption{Title, abstract and some excerpts  from a review on quantum annealing, discussing rigorous mathematical bounds for  errors and convergence times in different optimization cases.}
\label{Morita-08.pdf}
\end{figure}

\begin{figure}[!htbp]\centering

\includegraphics[width=17cm]{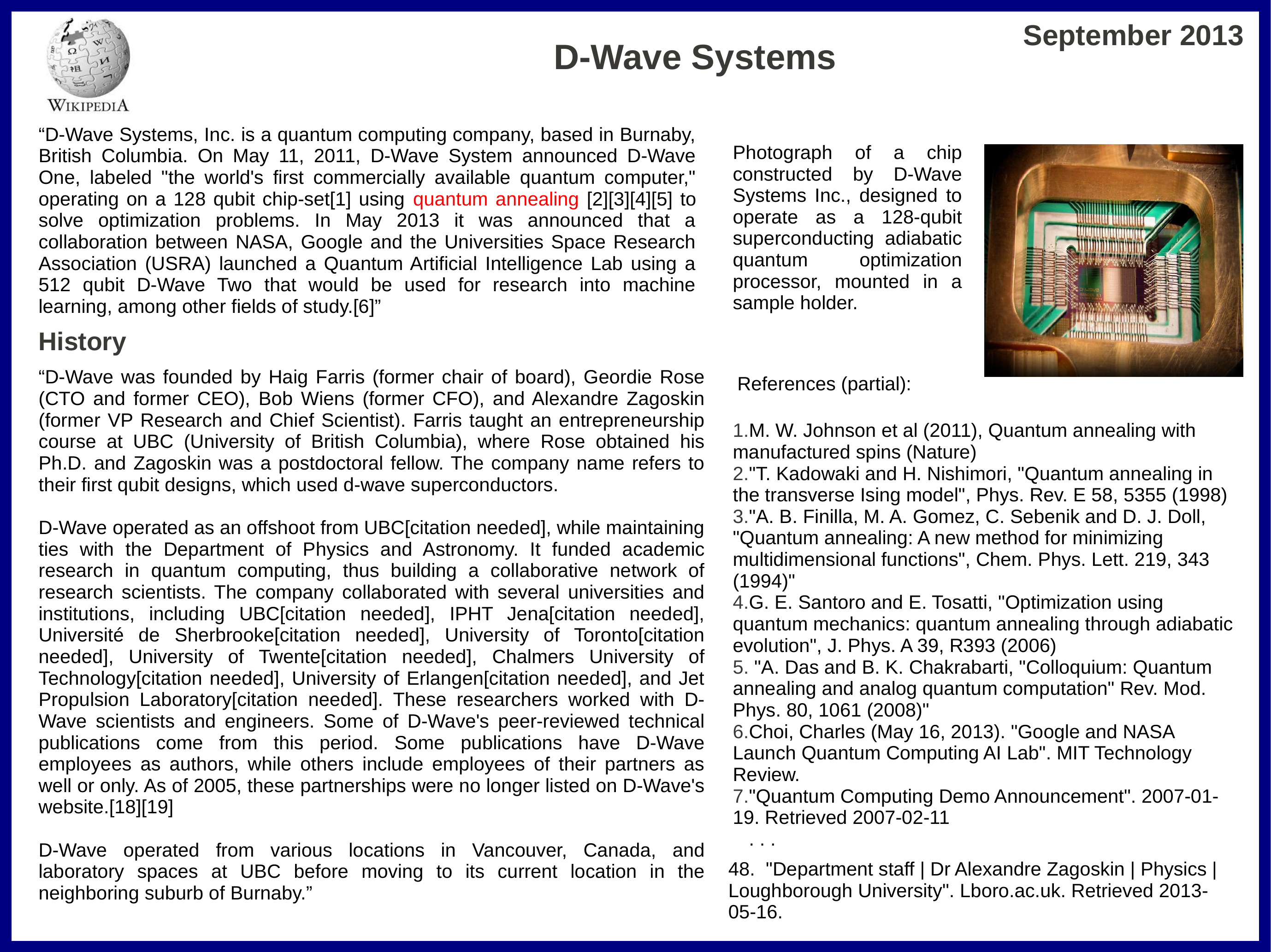}
\caption{From the Wikipedia entry on `D-Wave Systems Inc.' (as in September 2013;  website: http://en.wikipedia.org/wiki/D-Wave-Systems).}
\label{D-wave.pdf}
\end{figure}

\begin{figure}[!htbp]\centering
\includegraphics[width=14cm]{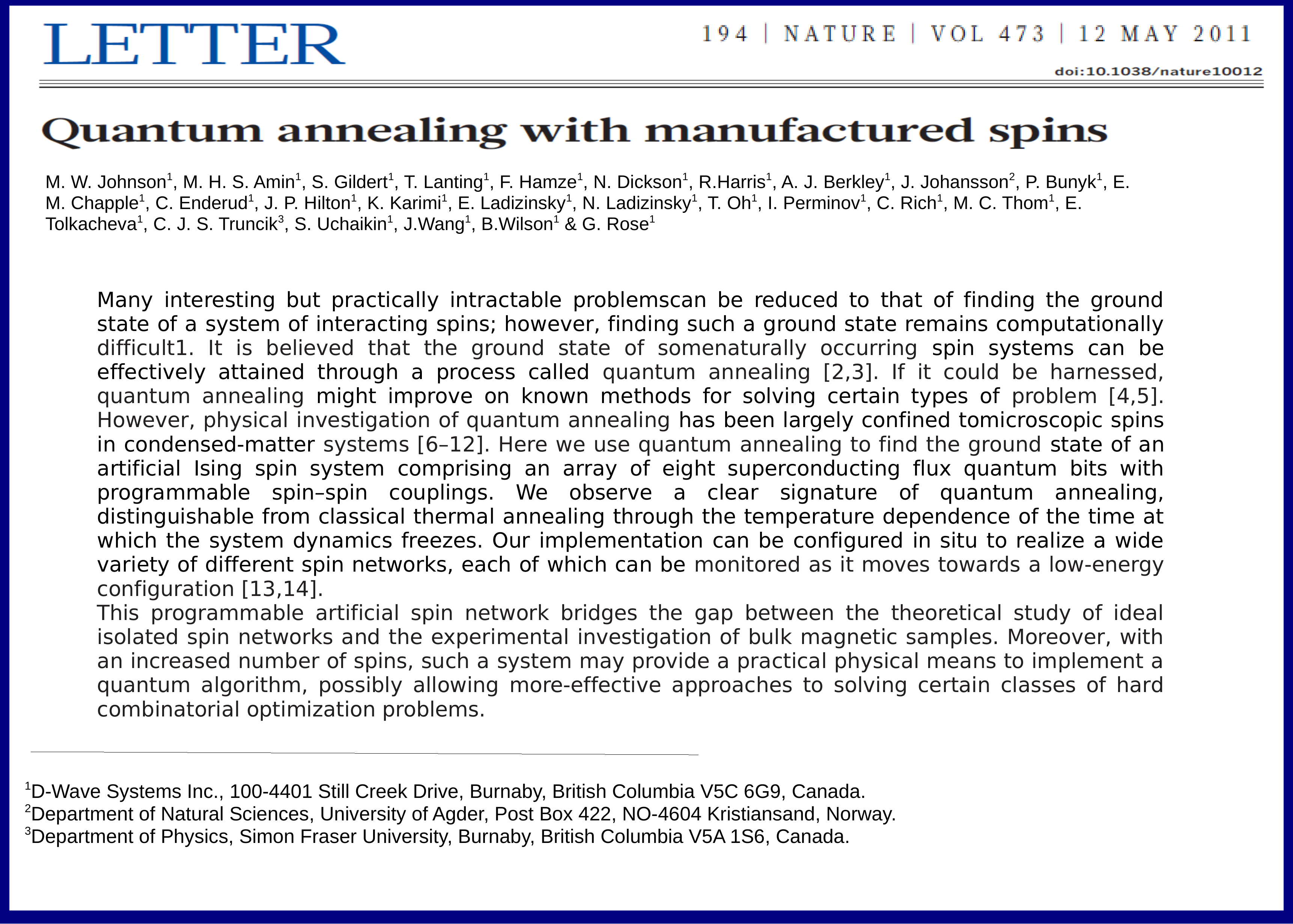}
\caption{Title and abstract of the first paper by D-Wave group giving the basic architecture of their quantum annealing precessor.}
\label{Johnson-12.pdf}
\end{figure}

\pagebreak

\begin{figure}[!htbp]\centering\vskip 2cm
\includegraphics[width=15cm]{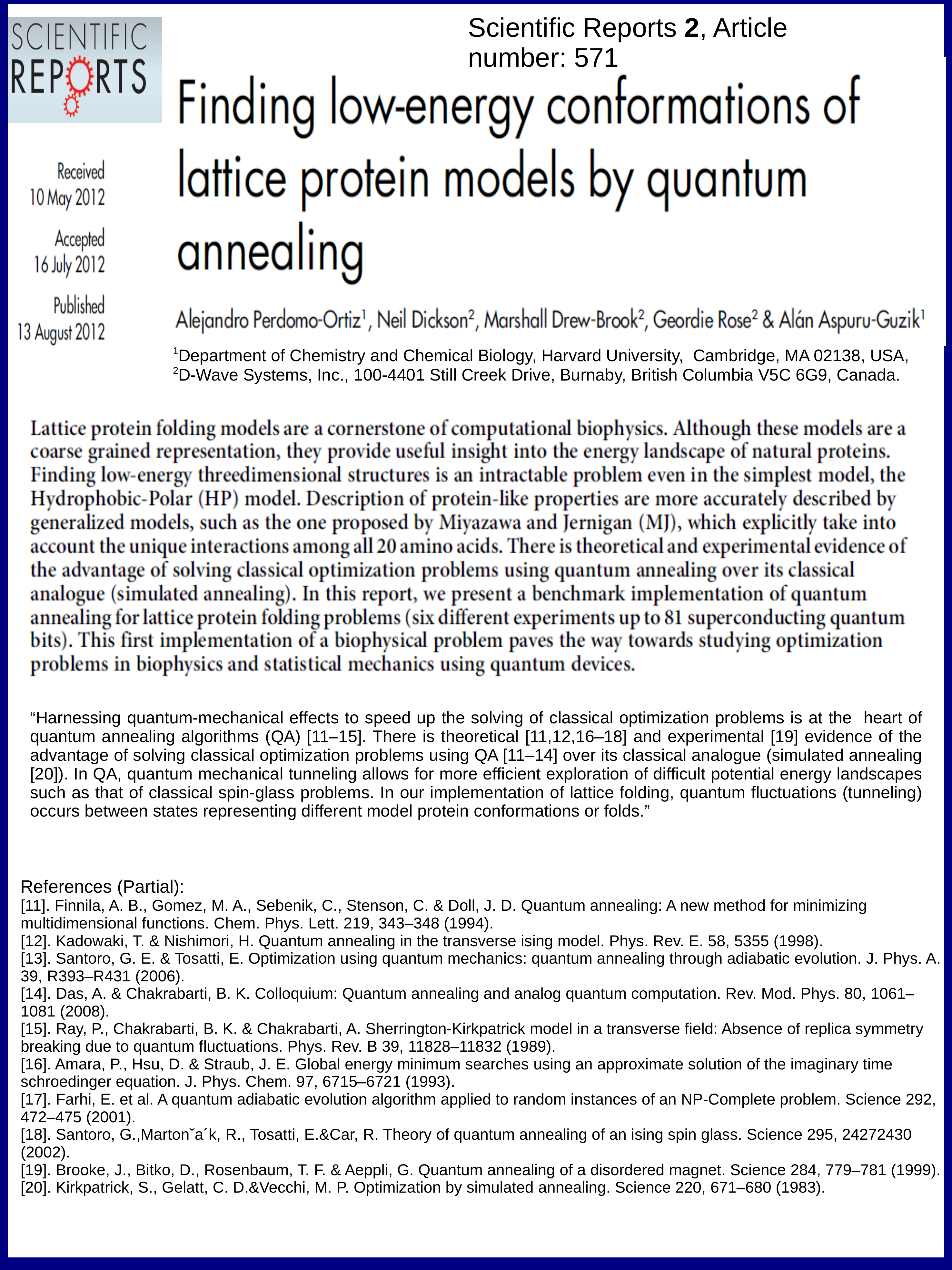}
\caption{Title, abstract and excerpts from the first major paper supporting the claim of D-Wave quantum computer used in searching the low energy conformations of lattice protein model, reported by the Harvard University group.}
\label{ortiz-12.pdf}
\end{figure}

\begin{figure}[!htbp]\centering

\includegraphics[width=16cm]{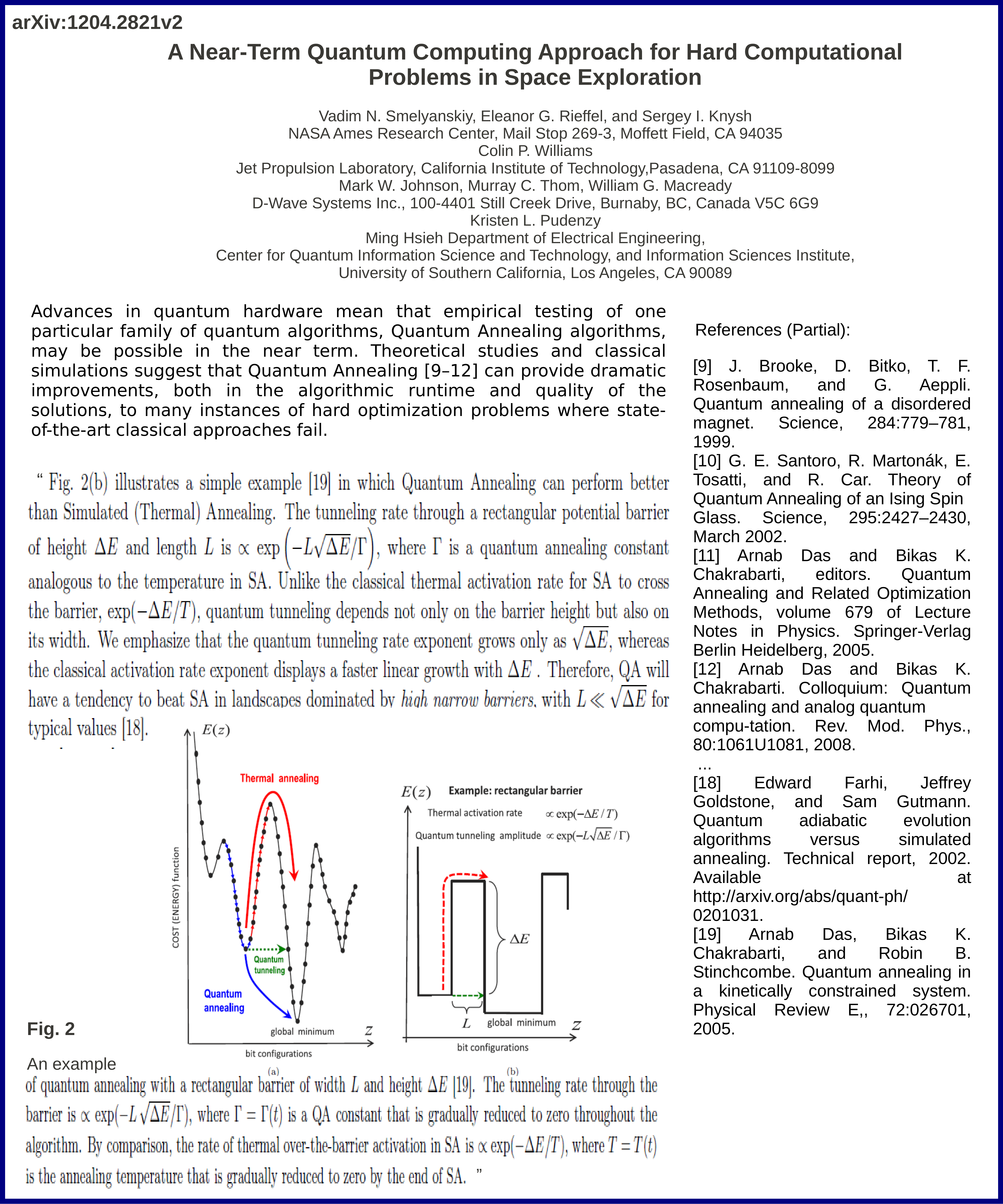}
\caption{Title, abstract and some excerpts of a  paper (from NASA Ames Research centre, Jet Propulsion Lab, CALTECH and University of Southern California) explaining  the basic principle of quantum annealing,  D-Wave computers and the possibilities of searching  solutions  of hard computational problems in space science and technology (website: http://arxiv.org/abs/1204.2821).}
\label{arXiv.pdf}
\end{figure}

\begin{figure}[!htbp]\centering

\includegraphics[width=14cm]{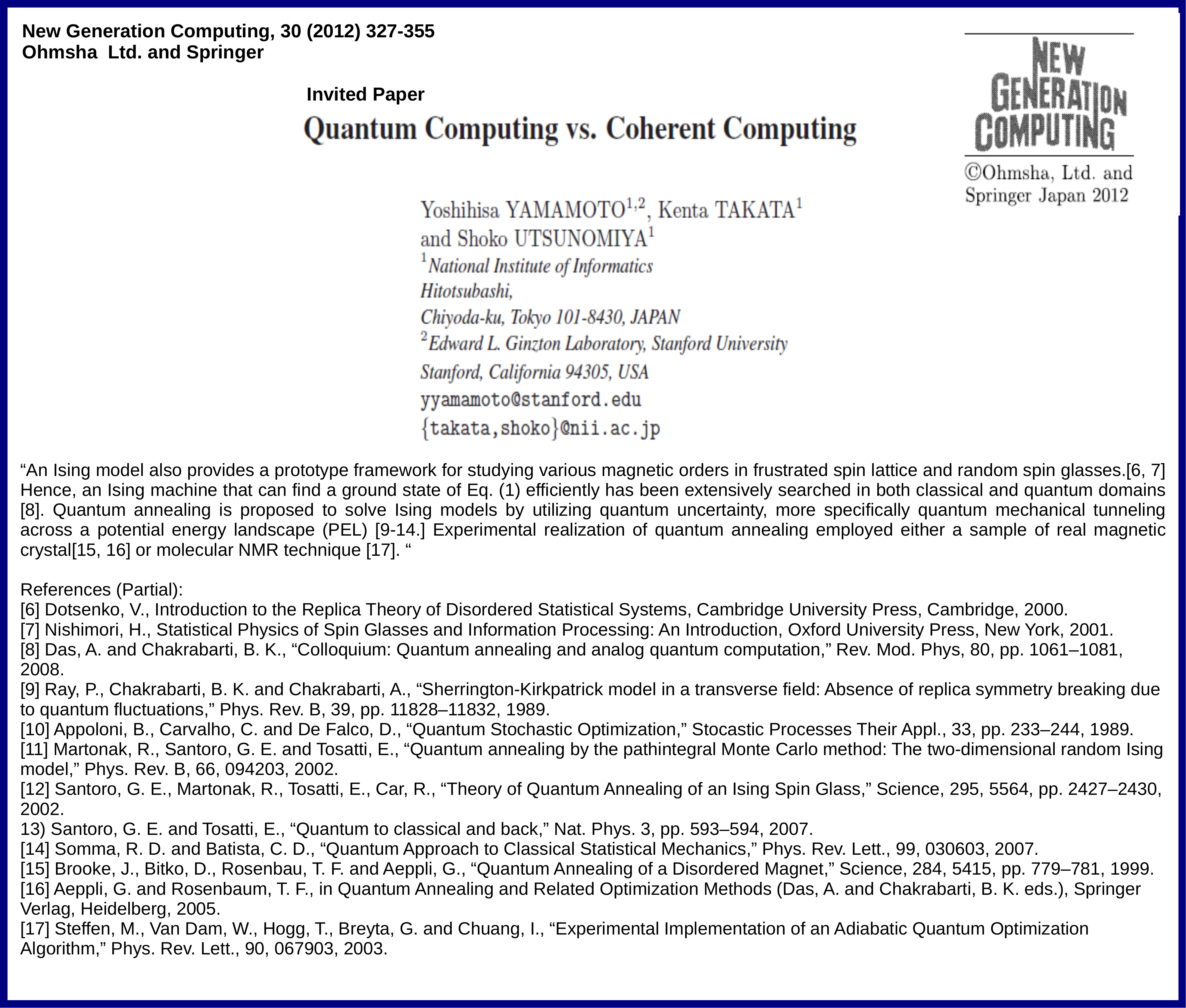}
\caption{Title  and some excerpts of a   paper (from National Institute of Informatics, Tokyo and Stanford University, California) comparing different quantum algorithms (including quantum annealing) in hardware platforms.}
\label{yamamoto.pdf}
\end{figure}
\begin{figure}[!htbp]\centering
\includegraphics[width=14cm]{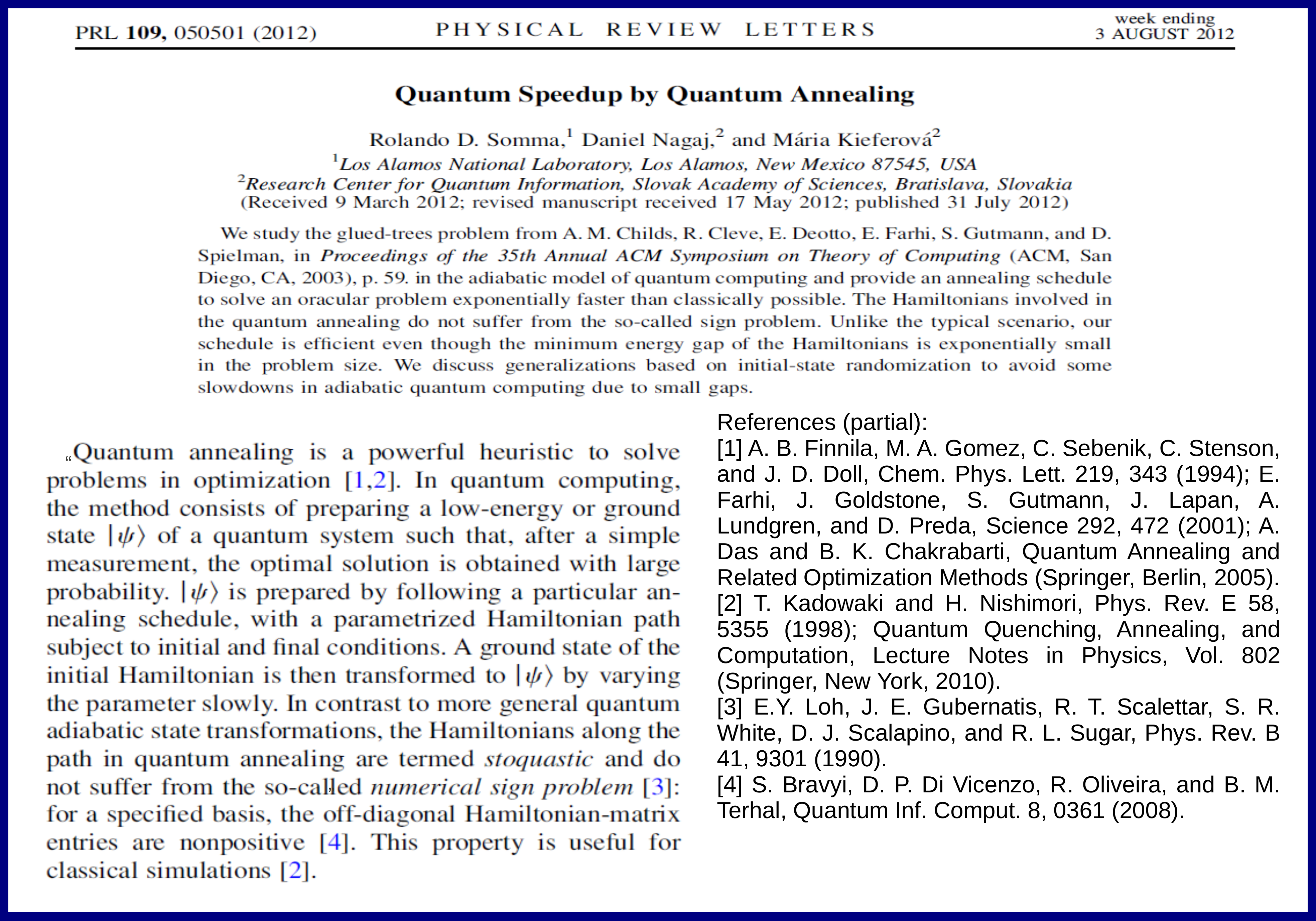}
\caption{Title, abstract  and some excerpts of a paper discussing possible speed-ups in quantum annealing computers.}
\label{Somma-12.pdf}
\end{figure}
\begin{figure}[!htbp]\centering
\includegraphics[width=14cm]{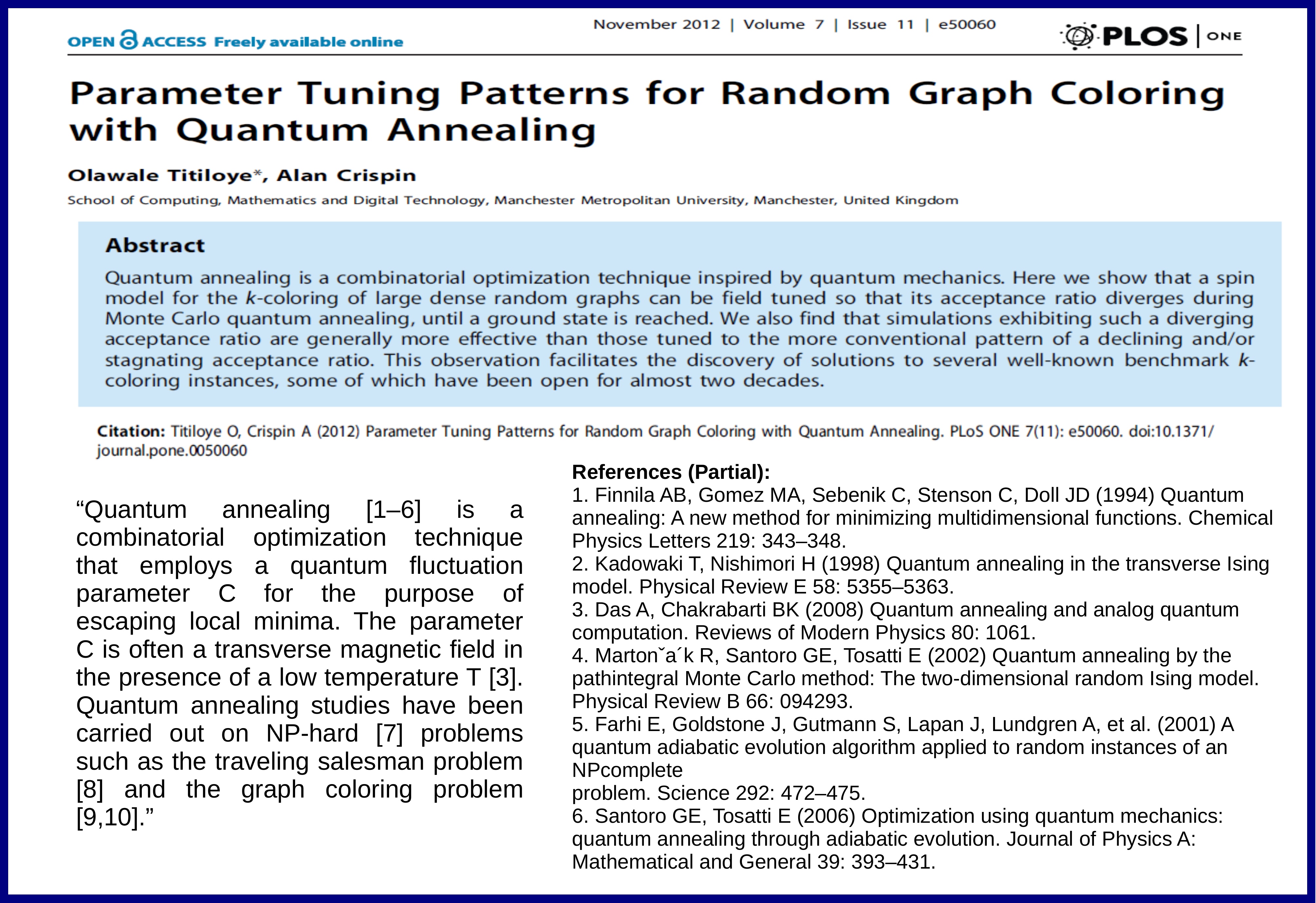}
\caption{Title, abstract  and some excerpts from a paper discussing successes of  quantum  annealing algorithms for graph-coloring problems.}
\label{titiloye-12.pdf}
\end{figure}
\pagebreak
\begin{figure}[!htbp]\centering
\includegraphics[width=14cm]{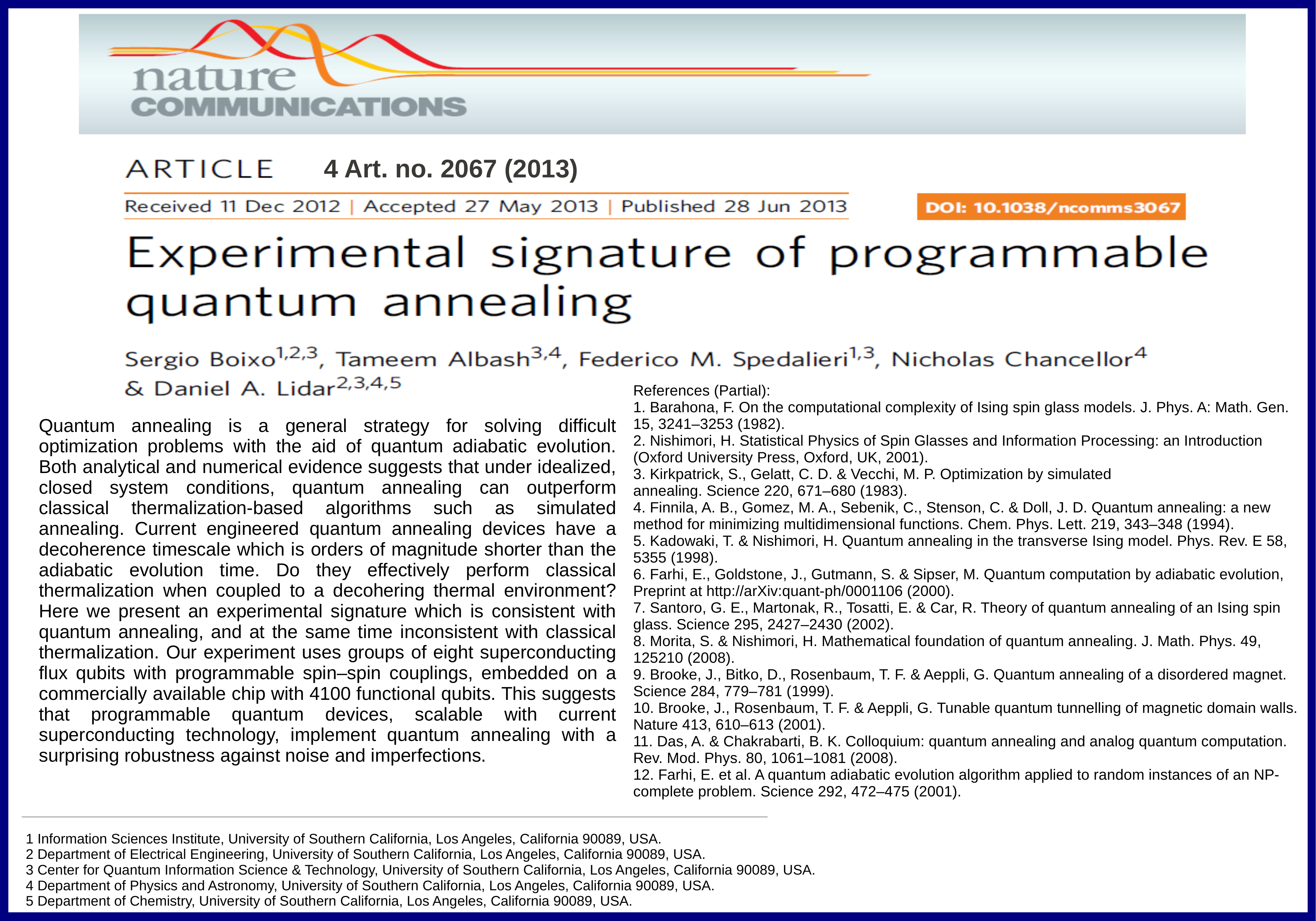}
\caption{Title and abstract of a paper reporting on quantum signatures in D-Wave machines and on their `surprising robustness against noise and imperfections'.}
\label{boixo-12.pdf}
\end{figure}

\begin{figure}[!htbp]\centering
\includegraphics[width=14cm]{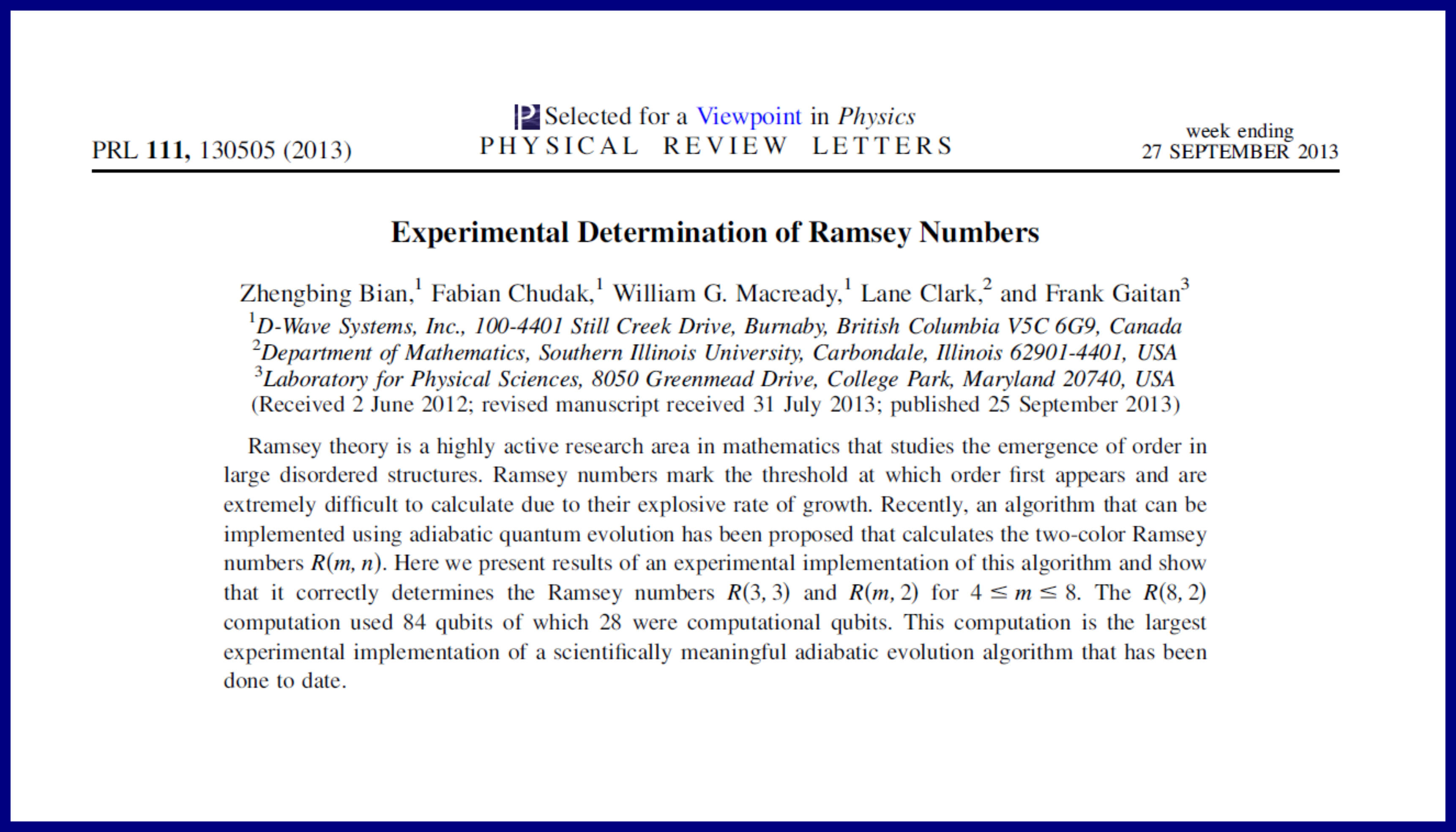}
\caption{Title and  abstract from ref. \cite{Bian-13} claiming a precise success (though very limited in scope) in a problem of party size calculation (with conflicting choices  among the  party members), using the D-wave computer.}
\label{bian-13.pdf}
\end{figure}

\begin{figure}[!htbp]\centering
\includegraphics[width=14cm]{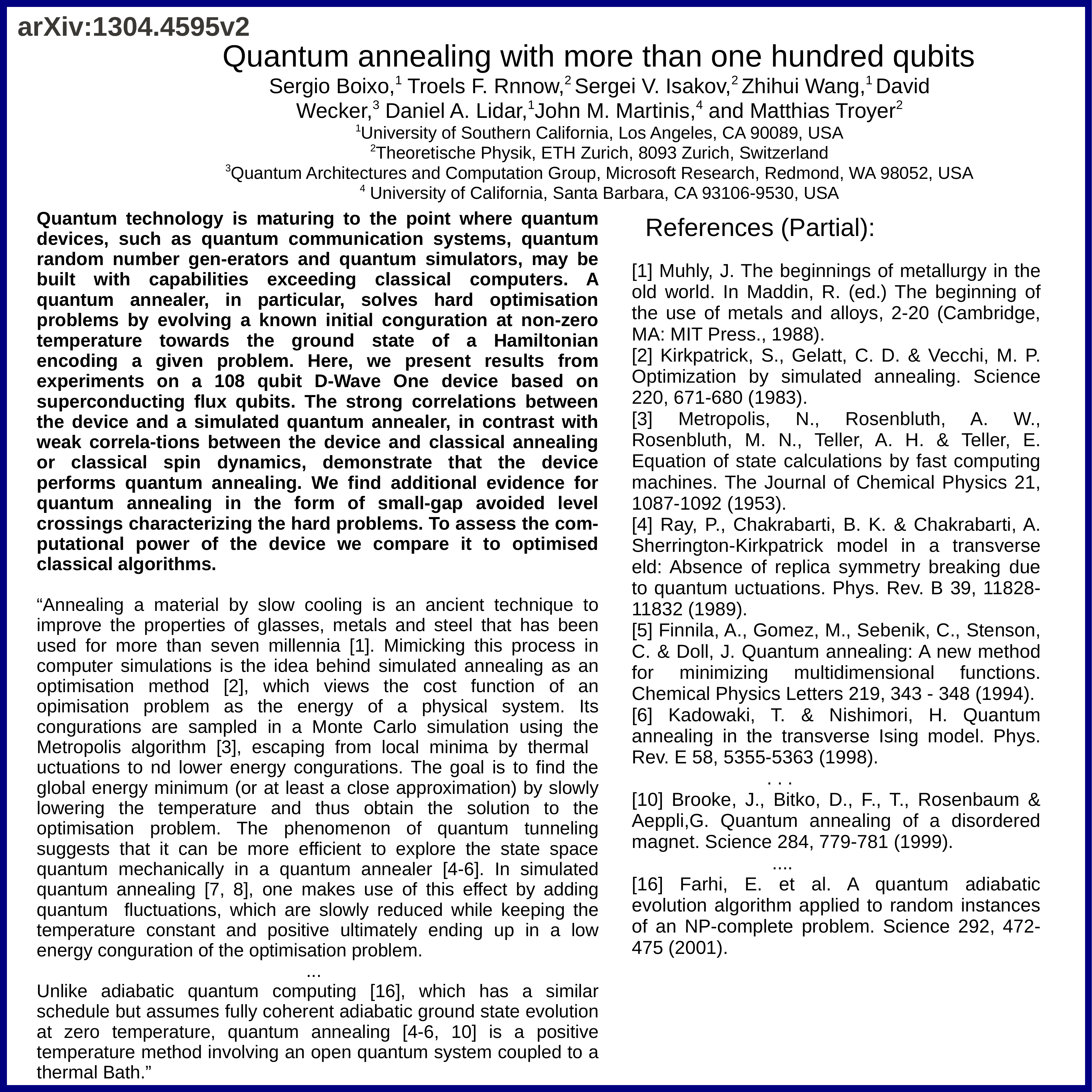}
\caption{Title, abstract and some excerpts from a paper by scientists from University of Southern California, University of  California, ETH Zurich and Microsoft Research, reporting on the precise quantum nature of the  performance of the D-Wave computer, compared with that of classical or  conventional  computers (website: http://arxiv.org/abs/1304.4595).}
\label{bioxo-13.pdf}
\end{figure}

\begin{figure}[!htbp]\centering
\includegraphics[width=16cm]{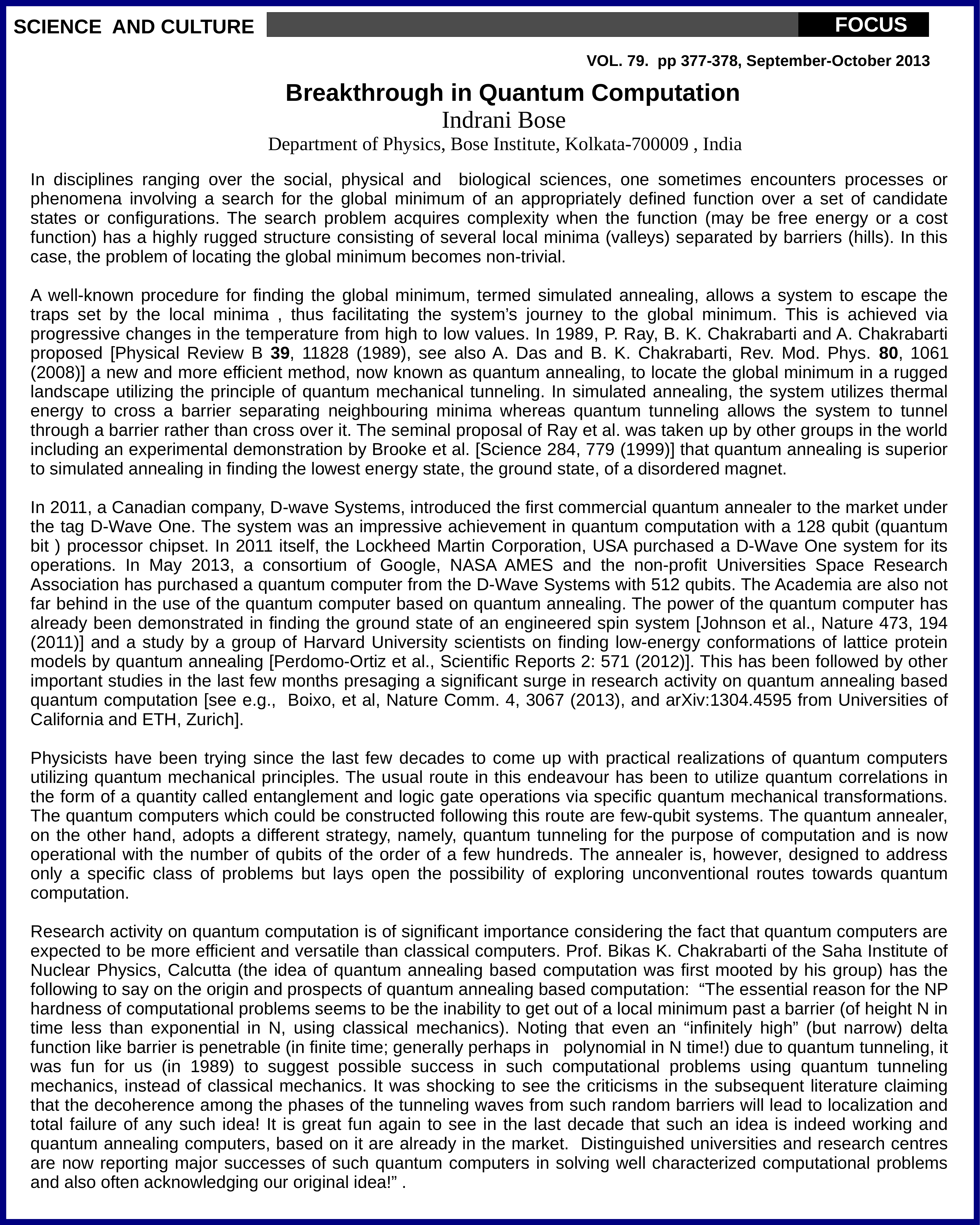}
\caption{A very recent news note by Bose \cite{bose-13} conveying the excitement.}
\label{bose-13.pdf}
\end{figure}
\pagebreak
\begin{figure}[!htbp]\centering
\includegraphics[width=14cm]{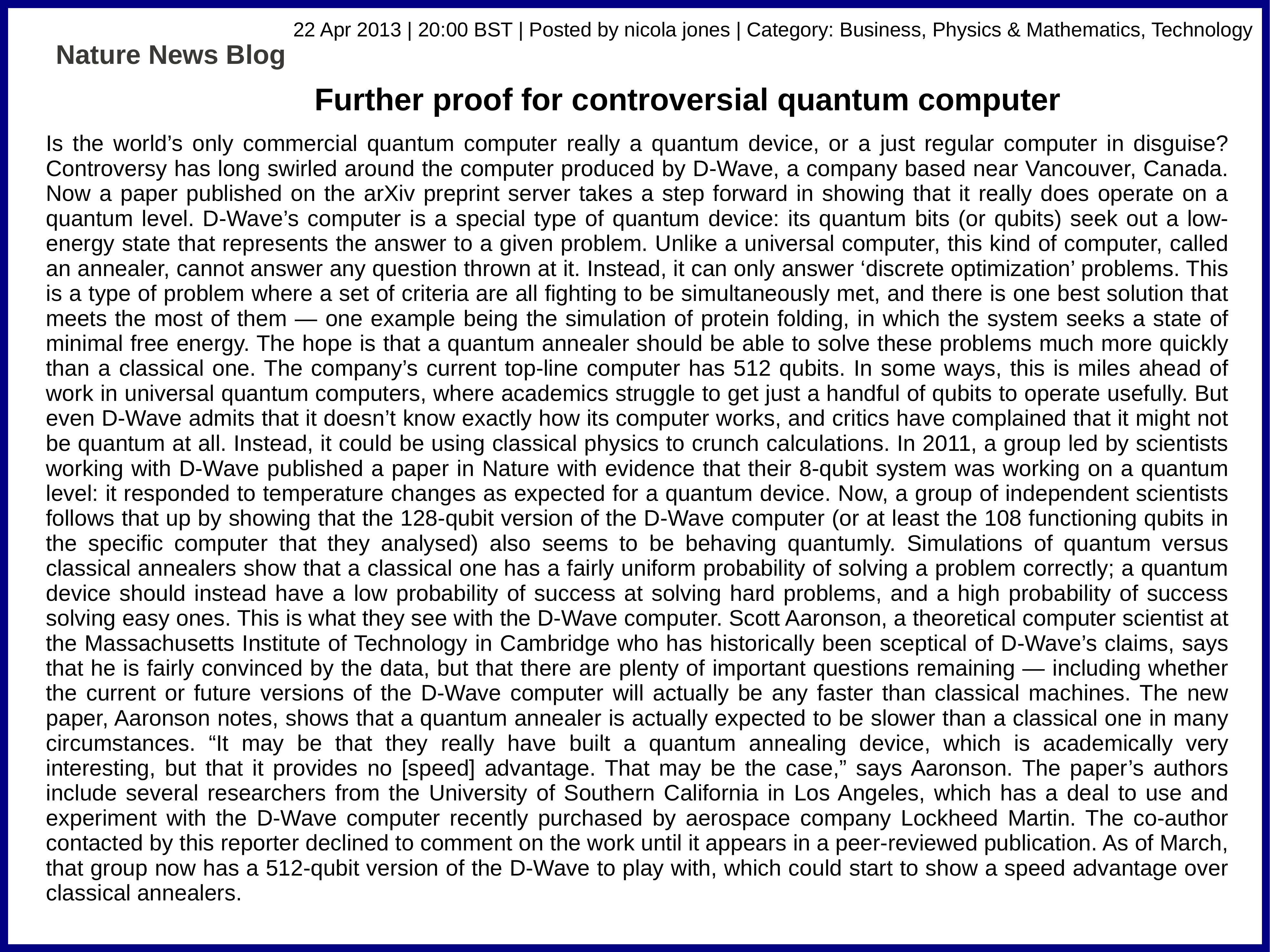}
\caption{Nature news blog regarding the recent reporting by scientists confirming the quantum nature of computation in D-Wave computers  (website: http://blogs.nature.com/news/2013/04/further-proof-for-controversial-quantum-computer.html).}
\label{nature.pdf}
\end{figure}


\section{Summary and conclusions}
Approximate solutions of computationally hard problems  were obtained more easily in Boltzmann-like machines which employ stochastic searches and (classical or thermal) annealing rather than employing sequential  search methods. It was noted  that for NP-hard problems,  the effective cost function landscape (in the solution state or configuration space) becomes extremely rugged as discussed in the text. Even with classical annealing tricks the probability  of escape from a local minimum to  another lower one separated by a barrier of height of order $N$ (the problem size) decreases as $\exp(-N)$, implying that the time to arrive at the solution in not bounded by any polynomial in $N$. Noting that the quantum tunneling probability  across such a barrier decreases   with width  of the barrier (becoming finite in the delta function barrier limit), Ray et al. \cite{ray-89}  proposed in 1989 that quantum tunneling might help solving NP-hard problems  in polynomial (in $N$) time! Subsequently, the researches by Finnila et al. \cite{finnila-94}, Kadowaki and Nishimori  \cite{kadowaki}, Brooke et al.  \cite{brooke-99} and Ferhi et al.  \cite{farhi-01} led to the robust development of quantum annealing technique, indicating  clearly the possible development of analog quantum computers with such tricks (cf.  Santoro and  Tosatti \cite{santoro-06}, Das and Chakrabarti \cite{das-08}). With the major breakthrough achieved by D-Wave computers \cite{johnson-11} (with indications \cite{Boixo-13} of search time $\sim\exp(N^\alpha)$; $\alpha<1$), a new era in  quantum computing has started: See e.g., \cite{seki-12,Titiloye-12,Yamamoto-12,Bapst,Boixo-nature-13}  etc.  as a few chosen examples only of the rapidly growing  publications  which (as shown in Figs. \ref{ortiz-12.pdf}, \ref{arXiv.pdf}, \ref{yamamoto.pdf}, \ref{boixo-12.pdf}, \ref{bioxo-13.pdf}) indicate also the role of the original papers in this  remarkable development. We note that the initial contributions by our Indian colleagues  had indeed been pioneering, though the follow-up researches and contributions have been rather slow.   

\section*{Acknowledgements}
This documentary  note was tentatively  planned during  the International Conference ``Diversity and Complexity: Realm of Today's Statistical Physics'' held in Saha Institute of Nuclear Physics, during 14-17 January, 2013 to commemorate sixty year of Prof. Bikas K. Chakrabarti. We are also grateful to Prof. Indrani Bose for exciting  us with her note on ``Breakthrough in Quantum Computation'' \cite{bose-13} (see Fig. \ref{bose-13.pdf}) and  to Dr. Arnab Das for many important comments and suggestions. We are thankful to   Dr. Masayuki Ohzeki for his comments and suggestions  and  to Dr. Diego de Falco for sending us some useful  documents.


\begin{thebibliography}{99}
\bibitem{ray-89}
P. Ray, B. K. Chakrabarti and A. Chakrabarti, \textit{Sherrington-Kirkpatrick model in a transverse field: Absence of replica symmetry breaking due to quantum fluctuations}, Physical Review B \textbf{39}, 11828 (1989).

\bibitem{Thirumalai-89}
D. Thirumalai, Q. Li and T. R. Kirkpatrick, \textit{Infinite-range Ising spin glass in a transverse field}, Journal of Physics A: Mathematical and General \textbf{22},  3339 (1989).


\bibitem{apolloni-89}
B. Apolloni, C. Carvalho, D. De Falco,\textit{Quantum stochastic optimization}, Stochastic Processes and their Applications, \textbf{33}, 233-244 (1989).

\bibitem{amara-93}
P. Amara, D. Hsu and J. E. Straub, \textit{Global energy minimum searches using an approximate solution of the imaginary time Schroedinger equation}, The Journal of Physical Chemistry \textbf{97}, 6715 (1993).

\bibitem{finnila-94}
A. B. Finnila, M. A. Gomez, C. Sebenik, C. Stenson and D. J. Doll, \textit{Quantum annealing: A new method for minimizing multidimensional functions}, Chemical Physics Letters \textbf{219}, 343 (1994).

\bibitem{chakrabarti-book}
B. K. Chakrabarti, A. Dutta and P. Sen,  \textit{Quantum Ising Phases and Transitions in Transverse Ising Models}, (see chapter 6 on Transverse Ising spin glass and random field systems, pp. 118-161), Springer, Heidelberg (1996).



\bibitem{kadowaki}
T. Kadowaki and H. Nishimori, \textit{Quantum annealing in the transverse Ising model}, Physical Review E \textbf{58}, 5355 (1998).

\bibitem{brooke-99}
J. Brooke, D. Bitko, T. F. Rosenbaum and G. Aeppli, \textit{Quantum annealing of a disordered magnet}, Science \textbf{284}, 779 (1999).

\bibitem{farhi-01}
E. Farhi, J. Goldstone, S. Gutmann, J. Lapan, A. Ludgren and D. Preda, \textit{A Quantum adiabatic evolution algorithm applied to random instances of an NP-Complete problem}, Science \textbf{292}, 472 (2001).

\bibitem{Santoro-02}
G. E. Santoro, R. Martonÿak, E. Tosatti and  R. Car, \textit{Theory of quantum annealing
of an Ising spin glass}, Science \textbf{295}, 2427 (2002).


\bibitem{das-05}
A. Das, B. K. Chakrabarti, and R. B. Stinchcombe, \textit{Quantum annealing in a kinetically constrained system}, Physical Review \textbf{72},  026701 (2005).


\bibitem{book-das}
 A. Das and B. K. Chakrabarti (Eds.), \textit{Quantum Annealing and Related Optimization Methods},  Springer, Heidelberg (2005).

\bibitem{santoro-06}
G. E. Santoro and E. Tosatti, \textit{Optimization using quantum mechanics: Quantum annealing through adiabatic evolution}, Journal of Physics A \textbf{39}, R393 (2006).


\bibitem{das-08}
A. Das and B. K. Chakrabarti, \textit{Quantum annealing and analog quantum computation}, Reviews of Modern Physics \textbf{80}, 1061 (2008).

\bibitem{morita-08}
S. Morita and H. Nishimori, \textit{Mathematical foundation of quantum annealing}, Journal of Mathematical Physics \textbf{49}, 125210 (2008).

\bibitem{Torres-08}
C. A.-Torres, D. M. Silevitch, G. Aeppli and T. F. Rosenbaum,  \textit{Quantum and classical glass transitions in $LiHo_{x}Y_{1-x}F_{4}$}, Physical Review Letters \textbf{101}, 057201 (2008).

\bibitem{book-Chandra}
 A. K. Chandra, A. Das and B. K. Chakrabarti (Eds.), \textit{Quantum Quenching, Annealing and Computation},  Springer, Heidelberg (2010).
\bibitem{johnson-11}
M. W. Johnson,  M. W. Johnson, M. H. S. Amin, S. Gildert,T. Lanting,F. Hamze, N. Dickson,R. Harris, A. J. Berkley, J. Johansson, P. Bunyk, E. M. Chapple, C. Enderud, J. P. Hilton, K. Karimi, E. Ladizinsky, N. Ladizinsky, T. Oh, I. Perminov, C. Rich, M. C. Thom, E. Tolkacheva, C. J. S. Truncik, S. Uchaikin, J. Wang, B. Wilson and  G. Rose,\textit{ Quantum annealing with manufactured spins}, Nature \textbf{473}, 194 (2011).

\bibitem{seki-12}
Y. Seki and H. Nishimori, \textit{Quantum annealing with antiferromagnetic fluctuations}, Physical Review E \textbf{85}, 051112 (2012).

\bibitem{smelyanskiy-12}
V. N. Smelyanskiy, E. G. Rieffel, S. I. Knysh, C. P. Williams, M. W. Johnson, M. C. Thom, and K. L. P. W. G. Macready, \textit{A near-term quantum computing approach for hard computational problems in space exploration}, arXiv:1204.2821 (2012).

\bibitem{Nagaj-12}
D. Nagaj, R. D. Somma, M. Kieferova, \textit{Quantum speed up by quantum annealing}, Physical Review Letters \textbf{109}, 050501 (2012).

\bibitem{Titiloye-12}
O. Titiloye, A. Crispin,  \textit{Parameter tuning patterns for random graph coloring with quantum annealing}, PLoS ONE \textbf{7(11)}, e50060 (2012).


\bibitem{Yamamoto-12}
Y. Yamamoto, K. Takata and S. Utsunomiya,  \textit{Quantum computing vs. coherent computing}, New Generation Computing \textbf{30},  327 (2012).


\bibitem{Ortiz-12}
A. P.-Ortiz, N. Dickson, M. Drew-Brook,	G. Rose	and A. A.-Guzik, \textit{Finding low-energy conformations of lattice protein models by quantum annealing}, Scientific Reports \textbf{2},  571 (2012).

\bibitem{Bapst}
V. Bapst, L. Foini, F. Krzakala, G. Semerjian and  F. Zamponi, \textit{The quantum adiabatic algorithm applied to random optimization problems: The quantum spin glass perspective}, Physics Reports \textbf{523}, 127 (2013).

\bibitem{Boixo-nature-13}
S. Boixo, T. Albash, F. M. Spedalieri, N. Chancellor and D.  A. Lidar, \textit{Experimental signature of programmable quantum annealing}, Nature Communications \textbf{4}, 2067 (2013).

\bibitem{Bian-13}
Z. Bian, F. Chudak, W. G. Macready, L. Clark and F. Gaitan,  \textit{Experimental determination of Ramsey numbers}, 
Physical Review  Letters  \textbf{111}, 130505 (2013).

\bibitem{Boixo-13}
S. Boixo, T. F. Rønnow, S. V. Isakov, Z. Wang, D. Wecker, D. A. Lidar, J. M. Martinis, M. Troyer, \textit{Quantum annealing with more than one hundred qubits},  	arXiv:1304.4595 (2013).

\bibitem{suzuki-13}
 S. Suzuki, J.-i. Inoue and  B. K. Chakrabarti, \textit{Quantum Ising Phases \& Transitions in Transverse Ising Models}, (see chapter 8 on Quantum annealing, pp. 225-290), Springer, Heidelberg (2013).


\bibitem{bose-13}
I. Bose,\textit{ Breakthrough in quantum computation}, Science and Culture \textbf{79}, 377 (2013). 

\bibitem{Dutta-book}
A. Dutta, G. Aeppli, B. K. Chakrabarti, U. Divakaran, T. Rosenbaum and D. Sen, \textit{Quantum Phase Transitions in Transverse Field Models: From Statical Physics to Quantum Information}, Cambridge University Press (forthcoming book, 2014). 

\bibitem{bkc-book}
 B. K. Chakrabarti, J.-i. Inoue, R. Tamura, and S. Tanaka, \textit{Quantum Spin Glasses, Annealing and Computation}, Cambridge University Press (forthcoming book, 2015). 

\end{thebibliography}
\end{document}